\documentclass[runningheads]{llncs}  

\usepackage{graphicx}
\usepackage{amsmath, amssymb}
\usepackage{xspace}
\usepackage{booktabs}
\usepackage{xcolor}
\usepackage{hyperref}
\usepackage{cleveref}
\usepackage{todonotes}  
\usepackage{caption}
\usepackage{multirow}
\usepackage{float}
\usepackage{arydshln}
\usepackage{tikz}
\usepackage{placeins}

\usetikzlibrary{arrows.meta,positioning,fit,calc}

\newcommand{\tri}{\textsc{TRIAGE}\xspace}
\newcommand{\tkg}{\textsc{TKG}\xspace}          
\newcommand{\tcs}{\ensuremath{\mathrm{TCS}}\xspace}          
\newcommand{\scr}{\ensuremath{\mathrm{SCR}}\xspace}          
\newcommand{\ec}{\ensuremath{\mathrm{EC}}\xspace}            
\newcommand{\xsc}{\ensuremath{\mathrm{CSC}}\xspace}          
\newcommand{\scomp}{\ensuremath{\mathrm{SC}}\xspace}         
\newcommand{\cvr}{\ensuremath{\mathrm{CVR}}\xspace}          
\newcommand{\csrlink}{\ensuremath{\mathrm{CSR_{link}}}\xspace} 
\newcommand{\ear}{\ensuremath{\mathrm{EAR}}\xspace}          
\newcommand{\tfi}{\ensuremath{\mathrm{TFI}}\xspace}          
\newcommand{\triti}{\ensuremath{\mathrm{TRIAGE\text{-}TI}}\xspace} 
\newcommand{\oc}{\ensuremath{\mathrm{OC}}\xspace}            
\newcommand{\dnr}{\ensuremath{\mathrm{DNR}}\xspace}          
\newcommand{\srr}{\ensuremath{\mathrm{SRR}}\xspace}          
\newcommand{\qgr}{\ensuremath{\mathrm{QGR}}\xspace}          
\newcommand{\erc}{\ensuremath{\mathrm{ERC}}\xspace}          
\newcommand{\erp}{\ensuremath{\mathrm{ERP}}\xspace}          
\newcommand{\rrs}{\ensuremath{\mathrm{RRS}}\xspace}          
\newcommand{\agr}{\ensuremath{\mathrm{AGR}}\xspace}          
\newcommand{\aur}{\ensuremath{\mathrm{AUR}}\xspace}          
\newcommand{\arf}{\ensuremath{\mathrm{ARF}}\xspace}          
\newcommand{\rhdm}{\ensuremath{\mathrm{RHD}}\xspace}          
\newcommand{\rpc}{\ensuremath{\mathrm{RPC}}\xspace}          
\newcommand{\gpc}{\ensuremath{\mathrm{GPC}}\xspace}          
\newcommand{\ctxcov}{\ensuremath{\mathrm{CC}}\xspace}        
\newcommand{\rfr}{\ensuremath{\mathrm{RFR}}\xspace}          
\newcommand{\kgimpact}{\ensuremath{\mathrm{KNI}}\xspace}    
\newcommand{\hhr}{\ensuremath{\mathrm{HHR}}\xspace}          
\newcommand{\mcorrect}{\ensuremath{\mu_{\mathrm{correct}}}}
\newcommand{\mcomplete}{\ensuremath{\mu_{\mathrm{complete}}}}
\newcommand{\mlc}{\ensuremath{\mu_{\mathrm{LC}}}}
\newcommand{\LCm}{\ensuremath{\mathrm{LC}}\xspace}     
\newcommand{\CORm}{\ensuremath{\mathrm{COR}}\xspace}   
\newcommand{\CMPm}{\ensuremath{\mathrm{CMP}}\xspace}   
\newcommand{\KG}{\ensuremath{\mathcal{K}}}                   
\newcommand{\refKG}{\ensuremath{\mathcal{K}^{*}}}            
\newcommand{\ont}{\ensuremath{\mathcal{O}}}                  
\newcommand{\sgrel}{\ensuremath{\mathrm{sim}}}               
\newcommand{\tkgterm}[1]{\textit{#1}}   

\begin{document}

\title{TRIAGE: Trustworthy Retrieval Instrumentation And Graph Evaluation}

\titlerunning{TRIAGE: Trustworthy Retrieval Instrumentation And Graph Evaluation}

\author{ Axel TahmasebiMoradi, Lucas Schott, Martin Royer } 
\institute{ IRT-SystemX \\
\email{
a.tahmasebimoradi@irt-systemx.fr,
lucas.schott@irt-systemx.fr,
martin.royer@irt-systemx.fr
}
}

\maketitle

\begin{abstract}

Knowledge graphs (KGs) that underpin Graph-based Retrieval-Augmented Generation (Graph-RAG) are increasingly built automatically by LLM-driven extraction rather than curated by experts.
Proper evaluation would require instrumenting all pertinent stages: extraction, graph construction, and inference, coherently enough to localize failures, so that a failure at one stage is not discovered as a wrong answer at the end.
We introduce \tri, a stage-aware instrumentation framework for automated, document-grounded graph-RAG that asks not only whether the underlying graph can be trusted but at what cost it can be queried.
\tri attaches stage-specific, independently interpretable metrics to three stages: the \tkgterm{KG Implementation} (triple confidence, source coverage, and schema and canonicalization checks), the \tkgterm{KG Validation} by expert (graph-level structural quality, with correctness and completeness computed only as offline calibration when a reference is available), and the \tkgterm{KG Usage} (retrieval coverage, faithfulness, and retrieval cost); the deployed metrics need no gold annotations, the gold-requiring ones serving only as offline calibration.
At usage time these metrics form a diagnostic chain of necessary conditions whose first broken link localizes the failure, and the diagnosis maps to the stage levers that can remedy it: extraction, graph and schema, or retrieval.
\tri is a theoretical framework with a proof of concept and a reproducible evaluation protocol.

\keywords{Knowledge Graphs \and Graph RAG \and Trustworthiness \and Trust Metrics \and Knowledge Graph Evaluation \and Automated Knowledge Graph Construction}
\end{abstract}

\section{Introduction}
\label{sec:intro}

\paragraph{Context and motivation.} Large language models (LLMs) have transformed information access by enabling fluent, query-driven generation over vast document collections.
Yet their most persistent failure mode remains hallucination: the generation of plausible but unsupported statements~\cite{ji2023survey,maynez2020faithfulness}.
Retrieval-Augmented Generation (RAG) was introduced precisely to mitigate this risk by grounding generation in externally retrieved evidence~\cite{lewis2020rag}.
However, when retrieval operates over flat vector indexes, it recovers semantically similar passages without capturing the relational structure that multi-hop or entity-centric questions demand~\cite{GraphRAG-Edge2024,chen2026pathrag}.
Knowledge graphs (KGs) offer a principled alternative: by representing factual knowledge as typed, directed triples $(s, p, o)$~\cite{hogan2021knowledge}, they make relationships explicit and support structured, traceable reasoning paths.
Graph-based RAG systems exploit this structure to improve retrieval precision and answer faithfulness~\cite{GraphRAG-Edge2024,LightRAG-Guo2024,chen2026pathrag}.
Yet the KG itself is increasingly built automatically by LLM-driven pipelines rather than curated by experts, introducing a new layer of uncertainty, even as we continue to judge such systems only by whether the final answer looks right.
The central question of this paper is therefore twofold: \emph{when} and \emph{how much} we can trust the graph that underpins retrieval, and at what \emph{cost}, in pathfinding time, retrieval latency, and compute, it can be queried.
A graph that is trustworthy but too costly to traverse is no more deployable than one that is cheap but wrong, so an end-to-end account of graph-RAG quality must speak to both.

\paragraph{The gap we are filling.} Recent work on trustworthy KG engineering, in particular the \tkg methodology~\cite{TKG-Amdouni2026}, provides a rigorous lifecycle framework covering construction, validation, deployment, and governance of KGs in safety-critical settings, and defines formal effectiveness metrics for correctness, completeness, and logical consistency.
However, \tkg and related approaches assume largely expert-driven knowledge elicitation and do not address the automated, LLM-driven extraction pipelines that underpin modern graph-RAG systems.
Conversely, graph-RAG systems such as GraphRAG~\cite{GraphRAG-Edge2024}, LightRAG~\cite{LightRAG-Guo2024}, and PathRAG~\cite{chen2026pathrag} are engineered for accuracy and efficiency, but provide no stage-specific trustworthiness instrumentation: failures in extraction, graph structure, or retrieval are invisible until they surface as wrong answers~\cite{eval-ragvsgraphrag2025,eval-reasoningbottleneck2026}.
Recent evaluation benchmarks have begun to characterize \emph{when} graph structure helps~\cite{eval-whentographs2025,eval-wildgraphbench2026}, and faithfulness-oriented work has studied whether answers are grounded in retrieved evidence~\cite{eval-finhallubench2026,eval-whatbreakskgrag2025}.
Yet none of these efforts instruments the pipeline at the level of individual stages (extraction, graph validation, and retrieval) with observable, ground-truth-free metrics that can localize failures and trigger principled remediation.
\tri fills this gap: it instruments the entire construction, validation, and use pipeline with gold-free, stage-localized trust and cost metrics, turning an opaque end-to-end failure into a diagnosis that pinpoints its stage and the levers that can fix it. 
We focus on automated, LLM-driven construction because that is how KGs are increasingly built, while evaluation has not kept up; the instrumentation, however, is not specific to it: the \tkgterm{KG Implementation} metrics apply to any extracted KG artifact, whether produced by LLM extraction or a conventional, non-AI pipeline.
\tri is deployment-first: its core trust and cost signals need no gold annotations, so they are computed on every live query, not only on a labeled benchmark.

\paragraph{Contributions.}
The contribution of \tri is not any single metric but a usable system: it locates where trust breaks across the pipeline, acts on each breach through the stage lever it implicates, surfaces which signals remain computable in a given deployment, and catches failures before the answer is generated.
Many of these metrics are individually simple, and several are adapted from prior work; what is new, and what makes them useful, is their organization into a stage-localized, deployment-ready instrument.
This paper makes the following contributions:
\begin{enumerate}
  \item \textbf{Framework.} Informed by the \tkg methodology~\cite{TKG-Amdouni2026}, we define \tri as a three-stage trustworthy engineering process for automated graph-RAG pipelines: \tkgterm{KG Implementation}, \tkgterm{KG Validation} (with expert review), and \tkgterm{KG Usage}.
    Rather than following \tkg prescriptively, \tri adapts its phase structure and makes independent design choices suited to automated extraction, each stage instrumented with stage-specific confidence metrics (\cref{sec:framework}).

  \item \textbf{Metrics.} We define a suite of stage-specific, independently interpretable metrics (overview in \cref{tab:metrics}), most of which are \emph{gold-free}: computable at deployment without any reference answer or graph.
    We further characterize them by two computability dimensions: what external reference each needs (none, an ontology, or a gold standard), and whether the extraction is observable to us (white-box) or only the final graph is given (black-box).
    At \tkgterm{KG Implementation}: Triple Confidence Score (\tcs), Source Coverage Ratio (\scr), Extraction Consistency (\ec), Cross-Source Corroboration (\xsc), Schema Compliance (\scomp), Constraint Violation Rate (\cvr), Canonicalization Success Rate (\csrlink), and Evidence Attribution Rate (\ear).
    At \tkgterm{KG Validation}: the graph-level, ground-truth-free Ontology Coverage (\oc), Dead Node Ratio (\dnr), Semantic Redundancy Rate (\srr), and Logical Consistency (\LCm), alongside correctness (\CORm) and completeness (\CMPm) imported from~\cite{TKG-Amdouni2026} for benchmark settings.
    At \tkgterm{KG Usage}: Query Grounding Rate (\qgr), Entity Retrieval Coverage (\erc), Entity Retrieval Precision (\erp), Reasoning Readiness Score (\rrs), Answer Grounding Rate (\agr), Answer Utilization Rate (\aur), Answer Reasoning Faithfulness (\arf), Retrieval Path Cost (\rpc), and Reasoning Hop Depth (\rhdm).
    Where gold paths or answers are available, Gold Path Coverage (\gpc) and Retrieval Path Precision/Recall (RPP/RPR) serve as offline checks, and the imported outcome metrics (Context Coverage, Reasoning Failure Rate, KG Net Impact, and Hard Hits Rate) as the dependent variables against which the gold-free metrics are validated (\cref{sec:metrics}).

  \item \textbf{Diagnosis, remediation, and validation.} The usage-stage metrics form a chain of necessary conditions whose first broken link localizes a failure to a single condition, and that diagnosis maps to the stage levers that can remedy it: \tkgterm{KG Implementation} metrics to extraction levers (prompt refinement, temperature reduction, re-chunking), \tkgterm{KG Validation} metrics to graph-and-schema levers (entity resolution, ontology fixes), and \tkgterm{KG Usage} metrics to retrieval levers, with aggregate metric drops triggering a new \tkgterm{KG Update} cycle that closes the loop toward continuous, monitored KG evolution (\cref{sec:conclusion}).
    We assume no universal cost order across these levers: which is cheapest is deployment-specific and is what our protocol measures (\cref{sec:experiments}).
    As a proof of concept we instantiate the usage metrics on a multi-hop KGQA probe and find that a capable reader can still answer from its parametric knowledge when retrieval misses the needed evidence, so the answer can look correct while retrieval has in fact failed; this motivates scoring retrieval before and independently of the answer, and a pre-specified, reproducible protocol then specifies the full validation (\cref{sec:experiments}).
\end{enumerate}

\paragraph{Paper structure.} \Cref{sec:related} surveys related work.
\Cref{sec:framework} presents the \tri architecture and the remediation map.
\Cref{sec:metrics} defines the metric suite.
\Cref{sec:experiments} presents the proof of concept and the evaluation protocol.
\Cref{sec:conclusion} concludes and outlines future work.

\section{Related Work}
\label{sec:related}

Trustworthiness in a graph-RAG pipeline is not a single property but a layered one: it must hold at the level of \emph{individual triples} (is each extracted fact correct and grounded?), at the level of the \emph{graph} they form (is the assembled KG complete, consistent, and well-structured?), and at the level of \emph{usage} (does retrieval surface the right subgraph, and does the generated answer faithfully reflect it?).
Prior work has addressed each level largely in isolation.
We organize this section accordingly.
We first review triple-level evaluation, distinguishing the assessment of extracted triples \emph{independently of the graph} (triple quality and trustworthiness) from their assessment \emph{as part of the graph} (knowledge graph completion).
We then review graph-based RAG methods, their evaluation protocols, and the related work on faithful and grounded reasoning over knowledge graphs.
Finally, we discuss trustworthy KG engineering methodologies and connect them to broader trustworthy AI principles, positioning \tri{} as a framework that instruments all three levels within a single, stage-aware methodology.

\subsection{Triple-centric related work: trustworthiness vs completion}

Knowledge graphs represent structured factual knowledge as directed labeled edges, commonly written as triples $(s,p,o)$, and a large body of work evaluates models and pipelines by how well they recover or validate such triples~\cite{bordes2013transe,trouillon2016complex,hoyt2022rankmetrics}.
We distinguish two evaluation objectives that differ in both their assumptions and their metrics, and that correspond to two successive moments in the construction of a KG: first assessing whether an extracted triple is trustworthy at all, and only then assessing how well a set of trusted triples completes a graph.

\textbf{Triple Quality / Trustworthiness.} The first question is whether a candidate triple, freshly proposed for insertion into a KG, is correct and well-formed.
Such triples may be produced by rule-based information extraction, neural relation extraction, Open Information Extraction (OpenIE), or, increasingly, LLM prompting.
Assessing them is a \emph{validation} problem: given a complete, already-formed triple $(s,p,o)$, decide whether it is true and admissible, a per-triple yes/no judgment rather than an ordering of alternatives.
This judgment must account for three distinct concerns: \emph{evidence grounding} (is the triple supported by the source text, or hallucinated?), \emph{schema compliance} (does it respect the ontology's predicate vocabulary and type constraints?), and \emph{logical consistency} (does it avoid contradicting other triples or ontology axioms?)~\cite{stanovsky2016oiebenchmark,bhardwaj2019carb,honovich2022true}.
Ensuring consistency with ontology constraints and type restrictions is central here, as violations (domain and range mismatches, type incompatibilities, or logical inconsistencies) often signal structural or semantic errors in extracted knowledge~\cite{nickel2015review,paulheim2017kgquality}.

\textbf{Knowledge Graph Completion (KGC).} Once a body of trusted triples exists, a second question arises: given the graph they form, which \emph{missing} edges can be inferred?
This is Knowledge Graph Completion, also known as link prediction, and unlike triple validation it is fundamentally a \emph{ranking} problem.
For a query of the form $(s,p,?)$ or $(?,p,o)$, where the relation is given and a missing \emph{entity} is sought, a model scores every candidate entity in the KG and ranks them, with success measured by how highly the correct entity is ranked~\cite{bordes2013transe,dettmers2018conve,sun2019rotate,hoyt2022rankmetrics}.
The distinction matters: validation asks ``is \emph{this} triple true?'' and returns a judgment about a single, complete triple; completion asks ``\emph{which} entity best completes this gap?'' and returns an ordering over many candidates.
Accordingly, KGC relies on rank-based metrics such as Mean Reciprocal Rank (MRR) and Hits@K, computed over both head- and tail-prediction tasks~\cite{bordes2013transe,trouillon2016complex,dettmers2018conve}.
Standard KGC evaluation further distinguishes \emph{raw} from \emph{filtered} settings, the latter removing other known true triples from the candidate set so that a model is not penalized for ranking a different correct answer above the target~\cite{bordes2013transe}.

A recurring methodological pitfall in KGC evaluation is benchmark contamination through inverse relations.
In early benchmarks such as FB15k, many relations have near-inverse counterparts: for instance, a \texttt{/film/directed\_by} edge and a \texttt{/film/director\_of} edge encode the same fact in opposite directions.
A model can then achieve high test scores not by learning meaningful relational structure, but by memorizing that one relation is the inverse of another: having seen $(A, \texttt{directed\_by}, B)$ at training time, predicting $(B, \texttt{director\_of}, A)$ at test time is trivial.
This inflates reported performance without reflecting genuine inference ability.
Corrected benchmarks such as FB15k-237 remove these redundant inverse and duplicate relations precisely so that the evaluation measures inference rather than memorization~\cite{dettmers2018conve,toutanova2015observed}.

\subsubsection{LLM-based triple extraction}

Before the adoption of LLMs, triple extraction relied on classical information-extraction methods (rule-based and dependency-pattern systems, and Open Information Extraction), which are deterministic and structurally consistent but brittle under linguistic variability and limited in implicit or cross-sentence reasoning~\cite{stanovsky2016oiebenchmark,banko2007textrunner}.
For the open-domain, paraphrase-rich setting that graph-RAG pipelines operate in, these approaches have been largely superseded by LLM-based extraction, which we focus on here.
LLM-based extraction uses generative models to map natural language directly into structured triples.
Compared to classical pipelines, LLMs are more effective at handling paraphrases, implicit relations, long-context dependencies, and some forms of multi-sentence reasoning~\cite{brown2020gpt3,wei2022chain}.
They are highly adaptable across domains with minimal feature engineering and can capture synonymy and nuanced language.
Crucially for trustworthiness, LLMs can be prompted to emit, alongside each triple, the source span from which it was derived, a natural-language justification or character offset into the input.
This provenance enables three things downstream: verification of the triple against its cited evidence, confidence estimation based on the strength of that evidence, and human audit of contested facts~\cite{honovich2022true,yao2023react}.
We operationalize exactly this signal later as the Evidence Attribution Rate (\ear, \cref{sec:metrics-trust}).

However, these advantages introduce new challenges.
LLMs may hallucinate plausible but unsupported facts~\cite{ji2023survey,maynez2020faithfulness}, and their outputs can vary depending on prompting and decoding strategies~\cite{wang2023selfconsistency}.
This leads to issues such as schema drift, non-deterministic formatting, and inconsistencies across runs.
The two standard prompting regimes trade one failure mode for another.
In zero-shot extraction, the model relies solely on instructions and input text, providing strong generalization but increasing the risk of hallucination and structural inconsistency~\cite{brown2020gpt3}.
Few-shot extraction augments prompts with demonstrations, improving precision and format control~\cite{min2022rethinking,zhao2021calibrate}, but the demonstrations themselves introduce \emph{demonstration bias}: the model tends to over-produce relation types and structures resembling the examples, while under-extracting valid triples that do not match the demonstrated patterns.
Neither regime is free of failure modes, which is precisely why dedicated trustworthiness metrics are needed rather than reliance on extraction accuracy alone.

The diversity of failure modes introduced by LLM-based extraction, such as hallucination, schema drift, demonstration bias, and output variability, exposes the limitations of evaluation metrics that rely solely on aggregate accuracy or exact matching.
In trustworthy graph construction, it is essential to quantify not only correctness but also confidence, coverage, and stability of extracted knowledge.
This motivates dedicated trustworthiness metrics such as the Triple Confidence Score (\tcs), which we operationalize here for extraction-level reliability within end-to-end graph-based RAG pipelines.

\subsection{Graph-based RAG}

Graph-based RAG~\cite{GraphRAG-Edge2024,LightRAG-Guo2024,chen2026pathrag} addresses a limitation of classical text RAG, namely its inability to capture structured dependencies across a document corpus.
Instead of indexing flat text chunks, these methods first construct an \emph{indexing graph} $G = (V, E)$ whose nodes $v \in V$ represent entities extracted from the documents and whose edges $(u, \rho, v) \in E$ encode their typed relations $\rho$.
At retrieval time, the graph structure is exploited to assemble a query-relevant subgraph that is verbalized as context for the generator.
Three landmark systems define the current design space, and the cleanest way to contrast them is by their \emph{retrieval primitive}, the granularity at which evidence is extracted from $G$ in response to a query $q$.

\textbf{GraphRAG}~\cite{GraphRAG-Edge2024} retrieves whole
\emph{community summaries}.
A hierarchical community detection (Leiden clustering) partitions the graph
into $\{C_1, \dots, C_m\}$, and an LLM produces a textual summary $s(C_i)$ for
each community.
The retrieved context for a query $q$ is the set of summaries deemed most
relevant to $q$:
\begin{equation*}
  R_{\mathrm{GraphRAG}}(q) \;=\; \big\{\, s(C_i) \,:\, C_i \in
  \operatorname{top-K}_{i}\, \mathrm{rel}(q, s(C_i)) \,\big\},
\end{equation*}
where $\mathrm{rel}(\cdot, \cdot)$ is an LLM-rated relevance score.
\emph{Pros:} the community primitive excels at global, dataset-spanning
summarization queries where the answer draws on a whole topic region.
\emph{Cons:} it produces verbose, redundant outputs whenever only a sub-region
of a community is pertinent, and the offline community-summarization step is
computationally heavy to build and to refresh as the graph changes.

\textbf{LightRAG}~\cite{LightRAG-Guo2024} retrieves the \emph{ego-networks}
of cosine-matched entity nodes.
An LLM extracts a keyword set $K_q$ from $q$ that is then matched against
entity embeddings via dense vector similarity:
\begin{equation*}
  V_q \;=\; \bigcup_{k \in K_q} \operatorname{top-N}_{\,v \in V}
  \cos\!\big(e(k),\, e(v)\big),
\end{equation*}
where $e(\cdot)$ denotes the embedding function, and the retrieved subgraph is
the 1-hop neighborhood of $V_q$.
\emph{Pros:} it is substantially lighter and faster than GraphRAG, with no
offline community-summarization cost, and its dual-level design adapts to both
specific and abstract queries.
\emph{Cons:} the ego-network of any query-related node still contains many edges
that are not on the reasoning chain linking the query's entities, introducing
structural noise that can mislead the generator.

\textbf{PathRAG}~\cite{chen2026pathrag} argues that the principal limitation of GraphRAG and LightRAG is the \emph{redundancy} of the retrieved subgraph rather than its insufficiency, and instead returns sparse multi-hop \emph{relational paths} between anchors.
The anchor set $V_q$ is computed by dense matching as in LightRAG; for each anchor a unit resource is propagated through the graph, decaying by $\alpha \in (0, 1)$ at every hop, and each path is scored by its average resource, with the top-$K$ paths forming the retrieved context.
\emph{Pros:} the path primitive is more compact than ego-networks and more directly aligned with the multi-step structure of the query, reducing the surface area over which the generator can hallucinate.
\emph{Cons:} path enumeration is expensive on dense graphs, where the number of candidate paths grows combinatorially; the resource-propagation scoring adds retrieval-time latency; and a sparse-path primitive can miss answers that require a fuller neighborhood rather than a single chain.

In short, the three systems trace a trade-off axis rather than a strict ranking: GraphRAG maximizes coverage at the cost of verbosity and offline expense, LightRAG minimizes cost at the risk of structural noise, and PathRAG maximizes precision at the cost of path-search latency.
Which point on this axis is appropriate depends on the query distribution and the available compute budget, a dependence that motivates measuring retrieval cost explicitly, as \tri does at \tkgterm{KG Usage} (\cref{sec:metrics-layer3}).

\subsubsection{Evaluation of graph-based RAG}

Because these systems target open-ended generation tasks for which no canonical ground-truth answer exists, the dominant evaluation protocol is \emph{LLM-as-judge} pairwise comparison along a small set of qualitative dimensions, introduced by GraphRAG and adopted, with minor variation, by LightRAG and PathRAG.
GraphRAG scores answers on Comprehensiveness, Diversity, and Empowerment, later extended with a Directness dimension~\cite{GraphRAG-Edge2024}; LightRAG uses Comprehensiveness, Diversity, and Empowerment together with an aggregate Overall judgment~\cite{LightRAG-Guo2024}; PathRAG follows the same protocol~\cite{chen2026pathrag}.
For each (query, baseline, candidate) triplet, an LLM selects a winner per dimension, and aggregate win-rates are reported.

This protocol is appropriate when no ground truth is available, but it has several limitations relevant to trustworthiness.
Judgments are stochastic and order-sensitive; the judge and the system under test often share the same underlying model family, raising self-preference concerns; and a win-rate is a coarse, black-box end-to-end signal.
The deeper problem is one of \emph{locus}: these protocols score only the generated \emph{answer}, never the retrieved \emph{subgraph}.
In plain terms, they check whether the final answer looks well-supported, but never check whether the retrieved subgraph actually contained the facts needed to answer the question.
When retrieval misses something, the LLM can quietly fill the gap from its own parametric memory and still produce a fluent, plausible answer, so a genuine retrieval failure looks identical to a genuine retrieval success.
Answer-level scores therefore cannot separate a sound reasoning path through the graph from a lucky guess by the model, which is exactly the distinction a trustworthy pipeline needs to make.

Beyond per-system judging, a first wave of recent benchmarks asks \emph{when} graph structure helps at all, but still scores only the end task~\cite{eval-ragvsgraphrag2025,eval-whentographs2025}.
\textbf{WildGraphBench}~\cite{eval-wildgraphbench2026} stresses the same task-level metrics under noisy, heterogeneous wild-source corpora, and \cite{eval-needgraphrag2026} adds a system-level dimension (accuracy, cost, latency, and stability) for graph-RAG versus dense RAG in agentic search.
Across this line, evaluation remains task-dependent and exposes no graph-aware signal: failures cannot be attributed to extraction, retrieval, or generation.

A second, faithfulness-oriented wave begins to open the black box, but its signals are still computed on the generated answer rather than on the retrieved subgraph.
\cite{eval-reasoningbottleneck2026} introduce a \emph{reasoning failure rate} alongside context coverage and token-F1 on multi-hop QA; \textbf{FinReflectKG-HalluBench}~\cite{eval-finhallubench2026}, a financial-QA benchmark, frames groundedness as binary hallucination detection, where an answer counts as grounded only if supported by both text snippets and KG triples; the tripartite RAG-Eval framework~\cite{eval-tripartite2025} aggregates query relevance, factual accuracy, coverage, coherence, and fluency into an overall confidence score; and \cite{eval-whatbreakskgrag2025} probes robustness under \emph{incomplete} KGs with Hits@Hard and the Hard Hits Rate.
These works diagnose whether an answer is grounded, but none instruments the structural adequacy of the retrieved subgraph itself, leaving the retrieval failure described above invisible.

\tri addresses this gap by attaching observable, ground-truth-free metrics directly at \tkgterm{KG Usage}: \qgr measures whether a query is on-topic for the KG; \erc, \erp, and \rrs characterize the structural quality of the retrieved subgraph independently of the downstream generation step; \agr, \aur, and \arf measure the alignment between the generated answer and the retrieved evidence; and \rpc measures the computational cost of retrieval itself.
These metrics complement the existing answer-level qualitative dimensions rather than replacing them: they isolate retrieval-stage failures from generation-stage failures with deterministic, structural signals that can trigger a \tkgterm{KG Update} cycle in a principled way.

\subsubsection{Faithful reasoning on KGs}

A complementary line of work studies \emph{faithful reasoning} over knowledge graphs: whether a system's answer can be traced, step by step, to explicit graph evidence rather than to parametric knowledge in model weights.
Graph-Constrained Reasoning (GCR)~\cite{GCR-Luo2024} operationalizes this directly by restricting the LLM's decoding to paths that exist in the KG, so that every reasoning step corresponds to a verifiable graph edge.
This substantially reduces hallucination on multi-hop questions, but it also exposes a new failure mode: when the KG itself is incomplete or incorrectly extracted, graph-constrained decoding fails silently, producing no answer, or a wrong one, with no signal indicating whether the fault lies in the graph or in the model.
A second, scale-related failure mode is equally important.
On large KGs, the number of candidate reasoning paths grows combinatorially with hop depth, so practical systems impose cutoffs: a maximum number of hops, a top-$K$ beam over paths, or a bounded expansion budget.
These cutoffs keep retrieval tractable, but they are double-edged: when the true reasoning path is longer than the hop limit, or is pruned by the beam, the correct answer becomes unreachable even though it is present in the graph.
This is a failure of \emph{traversal budget}, not of knowledge, and it is entirely invisible to answer-level metrics, directly motivating the Reasoning Hop Depth (\rhdm) and Retrieval Path Cost (\rpc) metrics introduced in \cref{sec:metrics-layer3}.

The same path primitive underlies earlier multi-hop KGQA systems~\cite{zhou2018interpretable,zhang2018variational}, which address faithfulness from the retrieval side: restricting context to sparse relational paths between query-relevant entities shrinks the surface over which hallucination can occur, the property PathRAG (above) brings to graph-RAG retrieval.
Evaluation on PathQuestion~\cite{zhou2018interpretable} and MetaQA~\cite{zhang2018variational} has shown that path coverage, whether the gold reasoning path lies in the retrieved subgraph, is a strong predictor of answer correctness on multi-hop questions.
This observation directly motivates the \tri Reasoning Readiness Score (\rrs) and Entity Retrieval Coverage (\erc), which operationalize path coverage as observable, pre-inference metrics that do not require gold answers.

Faithfulness has also been studied at the answer level, measuring whether generated statements are entailed by the retrieved evidence~\cite{honovich2022true,maynez2020faithfulness}.
As with the answer-level evaluation protocols above, such metrics require a generated answer and cannot separate insufficient retrieval from faulty reasoning; the \tri subgraph-level signals \rrs and \erc precede generation, enabling proactive triage rather than post-hoc auditing.

\subsection{Trustworthy KG engineering and trustworthy AI}

The \tkg methodology~\cite{TKG-Amdouni2026} is the closest work to our framework in spirit.
It proposes an end-to-end trustworthy engineering methodology for KG-based systems, structured along three complementary dimensions: a \tkgterm{methodology dimension} (construction phases per KG version), a \tkgterm{lifecycle dimension} (continuous evolution and updates), and a transverse \tkgterm{trustworthiness dimension} covering governance, provenance, and quality assessment across all phases.
Crucially, \tkg defines a formal suite of \tkgterm{effectiveness metrics}, correctness ($\mcorrect$), completeness ($\mcomplete$), logical consistency ($\mlc$), representativeness, and timeliness, which we directly import into the \tri \tkgterm{KG Validation} stage (see \cref{sec:metrics}).
However, \tkg was designed for expert-driven knowledge elicitation in safety-critical industrial settings and does not address automated, LLM-driven extraction pipelines or the \tkgterm{KG Usage} phase.
\tri is informed by its methodology and extends it to fill both gaps, while making independent design choices appropriate to automated graph-RAG.

Beyond \tkg, broader trustworthy AI frameworks, establishing principles of transparency, accountability, and robustness for AI systems, motivate the need for end-to-end, stage-aware evaluation rather than single-point quality checks~\cite{euhleg2019ethics,oecd2019ai,nist2023airmf,confianceai-gelin2024}.
Recent surveys on KG quality further consolidate the requirements for correctness, completeness, and consistency that \tri operationalizes for the graph-RAG setting~\cite{zaveri2016quality,farber2018linkeddataquality,mattioli2025kgmetrics,laudy2025hlif,paulheim2017kgquality}.
To the best of our knowledge, \tri is among the first frameworks to operationalize these principles end-to-end for automated, document-grounded graph-RAG pipelines, instrumenting extraction, validation, and usage within a single methodology.

\section{The TRIAGE Framework}
\label{sec:framework}

TRIAGE instruments a graph-RAG pipeline so that a failure can be read off its metrics rather than guessed at.
The payoff, developed in \cref{sec:remediation}, is twofold.
At usage time the metrics form a diagnostic chain of necessary conditions for a correct answer, and the first one that fails localizes the failure to a single condition.
That diagnosis then points to the stage levers that can remedy it: extraction, graph and schema, or retrieval.
The rest of this section sets up the three phases that carry these metrics, and \cref{sec:metrics} defines the metrics themselves; \cref{fig:worked} threads one query through all three.

\tri is informed by the \tkg engineering methodology~\cite{TKG-Amdouni2026} and adapts it to automated, document-grounded graph-RAG pipelines.
We take from \tkg its phase-structured view of KG engineering and its trustworthiness dimension, but we do not follow it prescriptively: the automated, LLM-driven setting calls for independent design choices that \tkg, conceived for expert-driven elicitation, does not address, notably the \tkgterm{KG Usage} phase and the remediation map developed in \cref{sec:remediation}.

Adapting the \tkg \tkgterm{methodology dimension}, \tri covers three phases of a single KG version cycle: \tkgterm{KG Implementation} (automated extraction and construction), \tkgterm{KG Validation} (expert review supported by ground-truth-free quality metrics), and \tkgterm{KG Usage} (graph-based inference).
Each phase produces both a \tkgterm{knowledge artifact} and a set of stage-specific confidence metrics.
The metrics do not flow between stages; rather, they are each independently observable quality signals that together provide an end-to-end trustworthiness picture of the pipeline.

\begin{figure}[t]
  \centering
  \begin{tikzpicture}[
    font=\footnotesize,
    >=Latex,
    node distance=5mm,
    proc/.style={draw=blue!60, fill=blue!5, rounded corners, align=center,
      inner sep=3pt, minimum width=30mm, minimum height=8mm},
    artifact/.style={draw=green!50!black, fill=green!6, rounded corners,
      align=center, inner sep=3pt, minimum width=24mm, minimum height=6.5mm},
    readout/.style={draw=gray!70, fill=gray!12, rounded corners, align=center,
      inner sep=3pt, minimum width=28mm, font=\scriptsize},
    update/.style={draw=orange!75!black, fill=orange!8, rounded corners, dashed,
      align=center, inner sep=3pt, minimum width=22mm, font=\scriptsize},
    tap/.style={gray!60, dashed, shorten >=1pt}
  ]
    \tikzset{
      lens/.pic={
        \draw[gray!75, line width=0.7pt, fill=white] (0,0) circle (1.9mm);
        \draw[gray!75, line width=1.3pt] (-1.35mm,-1.35mm) -- (-2.7mm,-2.7mm);
      }
    }
    \node[artifact] (docs) {Documents};
    \node[proc, below=of docs] (impl) {Phase 1: KG Implementation\\[-1pt]{\scriptsize extraction + grounding}};
    \node[artifact, below=of impl] (kgv) {KG$_v$};
    \node[proc, below=of kgv] (val) {Phase 2: KG Validation\\[-1pt]{\scriptsize expert review}};
    \node[artifact, below=of val] (kgval) {KG$_{validated}$};
    \node[proc, below=of kgval] (usage) {Phase 3: KG Usage\\[-1pt]{\scriptsize graph-based inference}};
    \node[artifact, below=of usage] (ans) {answer};
    \node[artifact, left=5mm of usage] (query) {query};
    \draw[->] (docs) -- (impl);
    \draw[->] (impl) -- (kgv);
    \draw[->] (kgv) -- (val);
    \draw[->] (val) -- (kgval);
    \draw[->] (kgval) -- (usage);
    \draw[->] (usage) -- (ans);
    \draw[->] (query) -- (usage);
    \node[readout, right=10mm of kgv] (mtrust) {Implementation metrics\\[-1pt]{\scriptsize see \cref{fig:phase1-pipeline}}};
    \node[readout, right=10mm of kgval] (mkgc) {Validation metrics\\[-1pt]{\scriptsize \oc\textsuperscript{\S},\ \dnr,\ \srr,\ \LCm\textsuperscript{\S}}};
    \node[readout, right=10mm of ans] (muse) {Usage metrics\\[-1pt]{\scriptsize see \cref{fig:phase3-pipeline}}};
    \draw[tap] (kgv.east) -- (mtrust.west);
    \draw[tap] (kgval.east) -- (mkgc.west);
    \draw[tap] (ans.east) -- (muse.west);
    \pic at ($(kgv.east)!0.5!(mtrust.west)$) {lens};
    \pic at ($(kgval.east)!0.5!(mkgc.west)$) {lens};
    \pic at ($(ans.east)!0.5!(muse.west)$) {lens};
    \node[update, right=6mm of mkgc] (update) {KG Update\\new cycle};
    \draw[->] (mtrust.east) to[out=0,in=180] (update.west);
    \draw[->] (mkgc.east) -- (update.west);
    \draw[->] (muse.east) to[out=0,in=180] (update.west);
    \draw[dashed,->] (update.north) .. controls +(0,16mm) and +(42mm,0) .. (impl.east);
  \end{tikzpicture}
  \caption{The \tri framework organized into three phases. Each phase produces
  distinct \tkgterm{knowledge artifacts} together with stage-specific confidence
  metrics (gray). Low aggregate metric values may trigger a new
  \tkgterm{KG Update} cycle; the stage of the failing metric points to the
  levers that can remedy it (\cref{sec:remediation}).}
  \label{fig:architecture}
\end{figure}
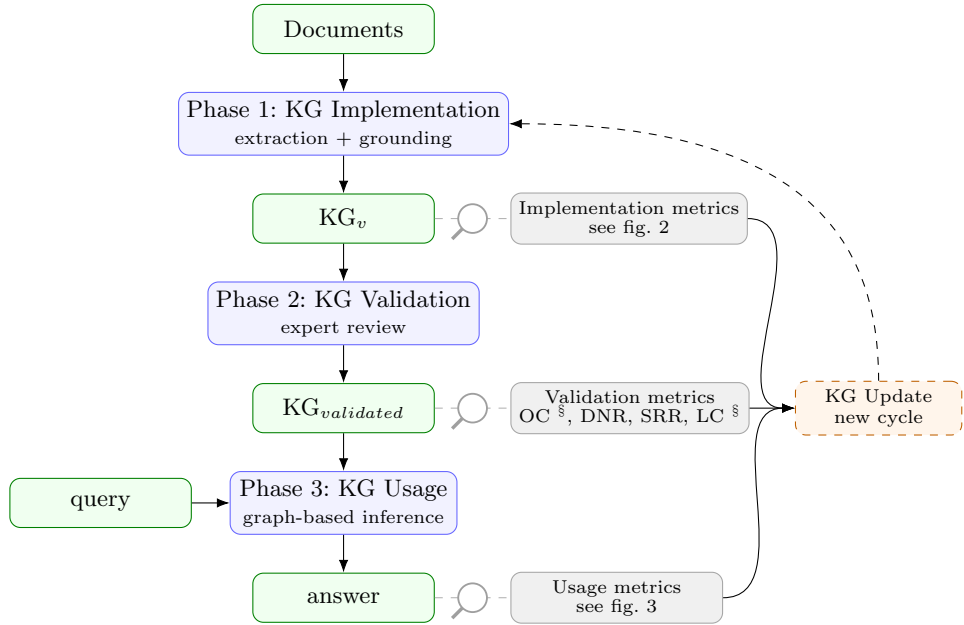

\Cref{fig:architecture} illustrates this structure.
The \tkg \tkgterm{trustworthiness dimension} operates transversally: aggregate metric results may trigger a \tkgterm{KG Update} cycle, and the stage at which a metric drops points to the levers that can remedy it (\cref{sec:remediation}).

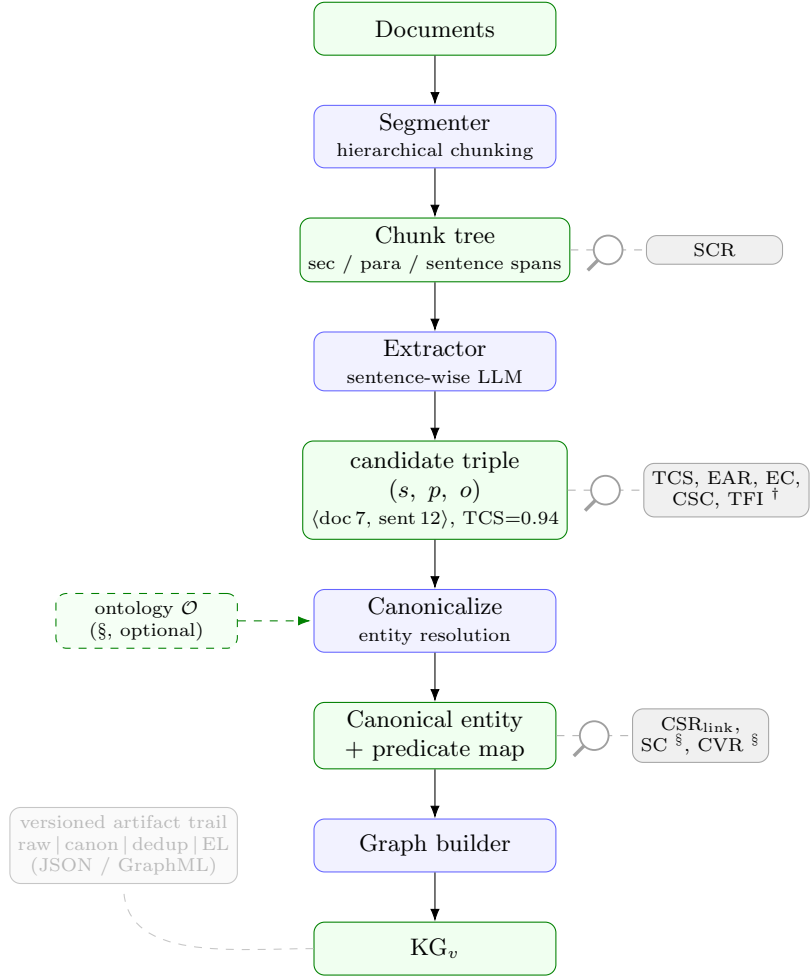
\begin{figure}[tbp]
  \centering
  \begin{tikzpicture}[
    font=\footnotesize,
    >=Latex,
    node distance=6.5mm,
    proc/.style={draw=blue!60, fill=blue!5, rounded corners, align=center,
      inner sep=3pt, minimum width=32mm, minimum height=7mm},
    artifact/.style={draw=green!50!black, fill=green!6, rounded corners,
      align=center, inner sep=3pt, minimum width=32mm, minimum height=7mm},
    card/.style={draw=green!50!black, fill=green!6, rounded corners,
      align=center, inner sep=3pt, minimum width=34mm, minimum height=13mm},
    onto/.style={draw=green!50!black, fill=green!4, rounded corners, dashed,
      align=center, inner sep=3pt, minimum width=24mm, font=\scriptsize},
    readout/.style={draw=gray!70, fill=gray!12, rounded corners, align=center,
      inner sep=3pt, minimum width=18mm, font=\scriptsize},
    faded/.style={draw=gray!45, fill=gray!4, rounded corners, align=center,
      inner sep=3pt, text=gray!55, minimum width=26mm, font=\scriptsize},
    tap/.style={gray!60, dashed, shorten >=1pt}
  ]
    \tikzset{
      lens/.pic={
        \draw[gray!75, line width=0.7pt, fill=white] (0,0) circle (1.9mm);
        \draw[gray!75, line width=1.3pt] (-1.35mm,-1.35mm) -- (-2.7mm,-2.7mm);
      }
    }
    \node[artifact] (docs) {Documents};
    \node[proc, below=of docs] (seg) {Segmenter\\[-1pt]{\scriptsize hierarchical chunking}};
    \node[artifact, below=of seg] (chunks) {Chunk tree\\[-1pt]{\scriptsize sec / para / sentence spans}};
    \node[proc, below=of chunks] (extr) {Extractor\\[-1pt]{\scriptsize sentence-wise LLM}};
    \node[card, below=of extr] (triple)
      {candidate triple\\ $(s,\ p,\ o)$\\[-1pt]{\scriptsize $\langle$doc\,7, sent\,12$\rangle$,\ $\tcs{=}0.94$}};
    \node[proc, below=of triple] (canon) {Canonicalize\\[-1pt]{\scriptsize entity resolution}};
    \node[artifact, below=of canon] (cmap) {Canonical entity\\ + predicate map};
    \node[proc, below=of cmap] (build) {Graph builder};
    \node[artifact, below=of build] (kgv) {KG$_v$};
    \draw[->] (docs) -- (seg);
    \draw[->] (seg) -- (chunks);
    \draw[->] (chunks) -- (extr);
    \draw[->] (extr) -- (triple);
    \draw[->] (triple) -- (canon);
    \draw[->] (canon) -- (cmap);
    \draw[->] (cmap) -- (build);
    \draw[->] (build) -- (kgv);
    \node[onto, left=10mm of canon] (ont) {ontology $\mathcal{O}$\\ (\S, optional)};
    \draw[tap,->,green!50!black] (ont.east) -- (canon.west);
    \node[readout, right=10mm of chunks] (rSCR) {\scr};
    \node[readout, right=10mm of triple] (rtrip)
      {\tcs,\ \ear,\ \ec,\\ \xsc,\ \tfi\textsuperscript{\dag}};
    \node[readout, right=10mm of cmap] (rcmap)
      {\csrlink,\\ \scomp\textsuperscript{\S},\ \cvr\textsuperscript{\S}};
    \draw[tap] (chunks.east) -- (rSCR.west);
    \draw[tap] (triple.east) -- (rtrip.west);
    \draw[tap] (cmap.east) -- (rcmap.west);
    \pic at ($(chunks.east)!0.5!(rSCR.west)$) {lens};
    \pic at ($(triple.east)!0.5!(rtrip.west)$) {lens};
    \pic at ($(cmap.east)!0.5!(rcmap.west)$) {lens};
    \node[faded, left=10mm of build] (trail)
      {versioned artifact trail\\ raw\,$|$\,canon\,$|$\,dedup\,$|$\,EL\\ (JSON / GraphML)};
    \draw[tap, gray!45] (kgv.west) to[out=180,in=-90] (trail.south);
  \end{tikzpicture}
  \caption{White-box instrumentation blueprint of the \tkgterm{KG Implementation}
  stage. The vertical spine is a representative construction pipeline (documents
  $\to$ segmentation $\to$ extraction $\to$ canonicalization $\to$ graph); each
  magnifier probe (right) taps one observable artifact: the chunk tree yields
  \scr; each candidate triple, carrying a provenance pointer and confidence,
  yields \tcs, \ear, \ec, \xsc, and \tfi{} (\dag, gold); and the canonical map
  yields \csrlink{} together with the schema checks \scomp{} and \cvr{} (\S,
  requiring the ontology $\mathcal{O}$, supplied as an optional side input). The
  persisted, versioned artifact trail (left) is what makes provenance auditable.
  Markers match \cref{tab:metrics}: \dag\ needs a gold reference, \S\ needs an
  ontology.}
  \label{fig:phase1-pipeline}
\end{figure}


\subsection{Phase 1: Document-to-Triple Extraction}
\label{sec:layer1}

To make the metrics concrete we describe a \emph{representative} extraction pipeline (\cref{fig:phase1-pipeline}); \tri instruments any pipeline of this shape rather than prescribing this one.
It turns document structure into candidate triples through an automated process of segmentation, extraction, grounding, canonicalization, and trust scoring, described step by step below.

\paragraph{Chunking and passage segmentation.}
The pipeline performs hierarchical segmentation before extraction:
\begin{enumerate}
  \item document into sections and subsections,
  \item subsection text into paragraphs,
  \item paragraph into sentence-like spans.
\end{enumerate}
The segmentation strategy determines which textual units are presented to the
extractor, directly influencing the coverage and recall of the resulting KG.

\paragraph{ML-based triple extraction.} Triple extraction is performed sentence-wise.
The extraction model follows an explicit instruction policy: emit only triples in \texttt{subject / predicate / object} format; preserve adverbs, modals, and negation in the predicate; and decompose conjunction-heavy clauses into multiple atomic triples.
This sentence-level granularity ensures that every extracted triple carries a precise provenance pointer and that confidence scores are computed at the finest possible resolution.
Each extracted triple is stored with provenance fields (document, section, subsection, paragraph index, sentence index, and sentence text), enabling passage-level traceability and downstream confidence scoring.

\paragraph{Canonicalization.}
Canonicalization keeps the graph coherent by ensuring that the same
real-world entity or relation is represented by a single node or predicate,
regardless of surface-form variation across source documents.
\tri applies it in one of two regimes, depending on whether a reference ontology
is available:
\begin{enumerate}
  \item \textbf{Ontology-grounded canonicalization} (when a schema is
    available): predicates and entities are normalized against the ontology;
    this enables schema checks and the Schema Compliance (\scomp) and Constraint
    Violation Rate (\cvr) metrics.
  \item \textbf{Resolution-based canonicalization} (when no suitable ontology
    exists): rule-based or embedding-based entity resolution, alias and
    coreference merging, and deduplication act as the fallback that keeps the KG
    connected and reduces dead nodes (\dnr) without a schema.
\end{enumerate}
This two-regime design reflects a broader property of \tri: it operates with or
without a reference ontology, degrading gracefully from schema-driven to
resolution-driven coherence.
The implications of ontology availability for \emph{measurement}, in particular which trustworthiness metrics remain computable without a schema,
are discussed in \cref{sec:metrics}.

\paragraph{Ontology grounding.} When an ontology is available, extracted triples are additionally aligned to it through predicate canonicalization against the target predicate inventory and optional schema checks evaluating predicate, domain, and range validity.
These checks produce a per-triple trust signal feeding \scomp and \cvr (\cref{sec:metrics-trust}).
The conceptual model is provided as input rather than constructed from scratch; how the relevant subset of the ontology is delineated in practice is an operational choice discussed in \cref{sec:layer2}.

\paragraph{Confidence scoring.} A Triple Confidence Score (\tcs) is assigned to each extracted triple as a proxy for extraction reliability.
The formal definition of \tcs and its role in the full metric suite are presented in \cref{sec:metrics}.
At the pipeline level, \tcs feeds document-level confidence aggregation and serves as an early-warning signal: triples whose \tcs indicates low confidence can be flagged for expert review during \tkgterm{KG Validation} (Phase~2) rather than silently entering the graph.

\paragraph{Post-processing and artifact persistence.} After extraction, the pipeline persists intermediate artifacts per stage (raw, canonicalized, deduplicated, and NER/EL-enriched) in JSON and GraphML format.
This versioned artifact trail directly supports the \tkg \tkgterm{trustworthiness dimension}: every triple in the final KG can be traced back to its source sentence and to the confidence score assigned at extraction time, which is what makes the Evidence Attribution Rate (\ear, \cref{sec:metrics-trust}) computable and the pipeline auditable.

\subsection{Phase 2: KG Validation, Expert Review and Effectiveness Assessment}
\label{sec:layer2}

Adapting the \tkg \tkgterm{KG Validation} phase~\cite{TKG-Amdouni2026}, the produced KG is reviewed by a domain expert before use.
\tri acts here as a \emph{methodological bridge}: it connects the automated extraction of Phase~1 to a human checkpoint, and makes that checkpoint actionable by surfacing interpretable, graph-level quality metrics as decision support.
Crucially, the core metrics of this phase require \emph{no ground-truth KG}, making them operational in any deployment setting; their formal definitions are given in \cref{sec:metrics-layer2}.
When a reference KG is available (e.g., in benchmark settings), correctness (\CORm) and completeness (\CMPm) from~\cite{TKG-Amdouni2026} can additionally be computed to calibrate the ground-truth-free metrics.

\begin{figure}[tbp]
  \centering
  \begin{tikzpicture}[
    font=\footnotesize,
    >=Latex,
    node distance=7mm,
    proc/.style={draw=blue!60, fill=blue!5, rounded corners, align=center,
      inner sep=3pt, minimum width=30mm, minimum height=7mm},
    artifact/.style={draw=green!50!black, fill=green!6, rounded corners,
      align=center, inner sep=3pt, minimum width=30mm, minimum height=7mm},
    readout/.style={draw=gray!70, fill=gray!12, rounded corners, align=center,
      inner sep=3pt, minimum width=20mm, font=\scriptsize},
    triage/.style={draw=orange!75!black, fill=orange!8, rounded corners, dashed,
      align=center, inner sep=3pt, minimum width=28mm, font=\scriptsize},
    faded/.style={draw=gray!45, fill=gray!4, rounded corners, align=center,
      inner sep=3pt, text=gray!55, minimum width=26mm, font=\scriptsize},
    tap/.style={gray!60, dashed, shorten >=1pt},
    gnode/.style={circle, draw=green!50!black, fill=green!55!black,
      minimum size=2.4mm, inner sep=0pt},
    onode/.style={circle, draw=gray!60, fill=gray!20, minimum size=2.4mm,
      inner sep=0pt}
  ]
    \tikzset{
      lens/.pic={
        \draw[gray!75, line width=0.7pt, fill=white] (0,0) circle (1.9mm);
        \draw[gray!75, line width=1.3pt] (-1.35mm,-1.35mm) -- (-2.7mm,-2.7mm);
      }
    }
    \node[artifact] (q) {query $q$};
    \node[proc, below=of q] (extr) {Entity Extraction\\[-1pt]{\scriptsize NER / noun-chunk}};
    \node[artifact, below=of extr] (Vq) {surface mentions $M_q$};
    \node[proc, below=of Vq] (ground) {Grounding $\gamma_q : M_q \to V \cup {\perp}$\\[-1pt]{\scriptsize entity linking}};
    \node[artifact, below=of ground] (N) {grounded query nodes $V_q^{\mathrm{g}}$};
    \node[proc, below=of N] (retr) {Retriever $R(\KG,V_q^{\mathrm{g}})$\\[-1pt]{\scriptsize ego / path / community}};
    \node[artifact, below=of retr, minimum width=38mm, minimum height=17mm] (S) {};
    \node[anchor=north, inner sep=2pt] at (S.north) {{\scriptsize retrieved subgraph $\KG_S$}};
    \node[proc, below=of S] (llm) {Verbalize \& Reader LLM};
    \node[artifact, below=of llm] (ans) {answer};
    \draw[->] (q) -- (extr);
    \draw[->] (extr) -- (Vq);
    \draw[->] (Vq) -- (ground);
    \draw[->] (ground) -- (N);
    \draw[->] (N) -- (retr);
    \draw[->] (retr) -- (S);
    \draw[->] (S) -- (llm);
    \draw[->] (llm) -- (ans);
    \node[gnode] (a1) at ([xshift=-12mm,yshift=-2.5mm]S.center) {};
    \node[onode] (mm) at ([xshift=-2mm,yshift=-4mm]S.center) {};
    \node[gnode] (a2) at ([xshift=10mm,yshift=-2.5mm]S.center) {};
    \node[onode] (xx) at ([xshift=4mm,yshift=0.5mm]S.center) {};
    \draw[green!50!black, line width=1.1pt] (a1) -- (mm) -- (a2);
    \draw[gray!55] (mm) -- (xx);
    \node[readout, right=12mm of N] (rQGR) {\qgr};
    \node[readout, right=12mm of retr] (rcost) {\rpc,\ \rhdm};
    \node[readout, right=12mm of S] (rS) {\erc,\ \erp,\ \rrs\\ \gpc$^{\dagger}$};
    \node[readout, right=12mm of ans] (rfaith) {\agr,\ \aur,\ \arf};
    \draw[tap] (N.east) -- (rQGR.west);
    \draw[tap] (retr.east) -- (rcost.west);
    \draw[tap] (S.east) -- (rS.west);
    \draw[tap] (ans.east) -- (rfaith.west);
    \pic at ($(N.east)!0.5!(rQGR.west)$) {lens};
    \pic at ($(retr.east)!0.5!(rcost.west)$) {lens};
    \pic at ($(S.east)!0.5!(rS.west)$) {lens};
    \pic at ($(ans.east)!0.5!(rfaith.west)$) {lens};
    \node[triage, left=12mm of llm] (tri)
      {pre-inference triage\\ \erc\ or\ \rrs $<\theta \to$ reroute / warn};
    \draw[tap,->,orange!75!black] (S.west) to[out=180,in=90] (tri.north);
    \node[faded, left=12mm of ans] (bb)
      {black-box fallback: LLM-as-judge\\ (uses $q$, answer only)};
    \draw[tap, gray!45] (ans.west) -- (bb.east);
    \pic at ($(ans.west)!0.5!(bb.east)$) {lens};
  \end{tikzpicture}
  \caption{White-box instrumentation blueprint of the \tkgterm{KG Usage} stage.
  The vertical spine is the inference path; each magnifier probe (right) taps one
  \emph{observable} artifact and lists the metrics computable from it: the surface mentions \(M_q\) and grounded query nodes \(V_q^g\) yield \qgr; the retrieved subgraph $\KG_S$ yields \erc, \erp, and \rrs
  (and \gpc when gold paths are available, $\dagger$); the retrieval trace yields
  the cost metrics \rpc and \rhdm; and the answer, compared against $\KG_S$, yields
  \agr, \aur, and \arf (of these \arf and \agr are the chain's answer-side links, \aur their utilization complement). The chain heads \qgr, \erc, \rrs are read \emph{before}
  the reader runs, enabling pre-inference triage (left). A black-box evaluator
  (bottom left) sees only the query and the answer, so it cannot localize a
  retrieval failure. Inside $\KG_S$, filled nodes are anchors and the bold edge is
  the reasoning path connecting them.}
  \label{fig:phase3-pipeline}
\end{figure}
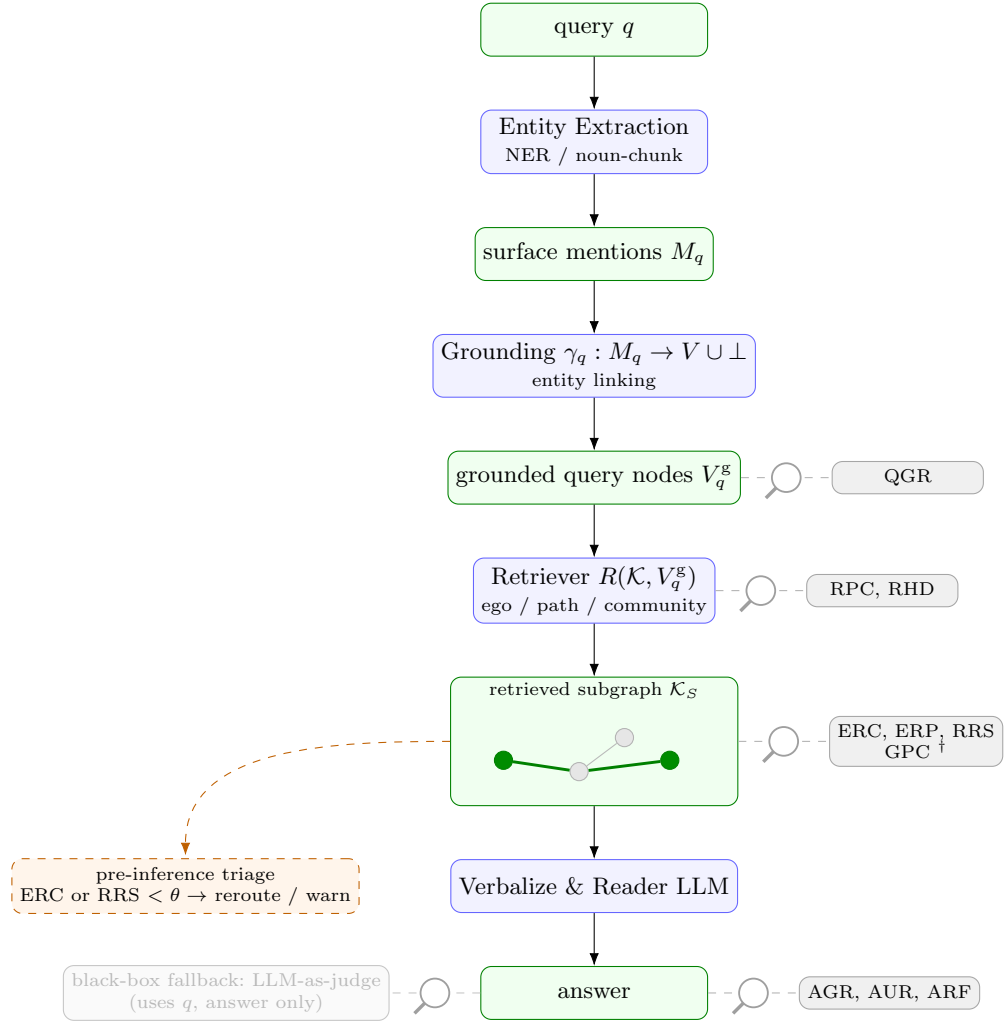

\subsection{Phase 3: KG Usage, Graph-Based Inference}
\label{sec:layer3}

The \tkgterm{KG Usage} phase is identified in \tkg~\cite{TKG-Amdouni2026} as a gap in prior KG engineering methodologies.
\tri operationalizes it for graph-RAG systems via graph-based retrieval and inference: a query is grounded to the KG, a query-relevant subgraph is retrieved, and that subgraph conditions answer generation.
Graph-constrained decoding~\cite{GCR-Luo2024} is one such inference method, but the usage-specific metrics introduced here are agnostic to the choice of retriever and reader; they characterize the structural adequacy of the retrieved subgraph, the faithfulness of the generated answer, and the computational cost of retrieval (\cref{sec:metrics-layer3}).

\paragraph{Inference pipeline.} Given a validated KG and a natural language query $q$, the \tri inference layer proceeds in three steps (\cref{fig:phase3-pipeline}).
First, query grounding maps $q$ to a set of anchor entities in the KG via entity linking and embedding-based similarity.
Second, subgraph retrieval extracts a query-relevant subgraph from the full KG; the specific retrieval strategy (ego-network, path-based, or community-based) is configurable and does not affect metric computation.
Third, the retrieved subgraph is verbalized and supplied to an LLM to generate the final answer, for example via graph-constrained decoding~\cite{GCR-Luo2024}.

\paragraph{Pre-inference triage.} A key design principle of the \tri \tkgterm{KG Usage} phase is that structural quality signals are computed \emph{before} generation, enabling proactive triage rather than post-hoc diagnosis.
\erc measures whether the entities required to answer $q$ are present in the retrieved subgraph, and \rrs measures whether those entities are connected by a valid reasoning path within it.
Queries for which \erc or \rrs fall below a configurable threshold can be flagged before the LLM is invoked, routed to a fallback retrieval strategy or returned with a low-confidence warning, avoiding the silent failure mode in which the LLM compensates with parametric knowledge and produces a plausible but unsupported answer.

\begin{figure}[t]
\centering
\fbox{\begin{minipage}{0.95\linewidth}
\small
\textbf{Worked example (end to end, on MetaQA / WikiMovies; illustrative values).}
\emph{Query (2-hop):} ``Did the director of \textit{Inception} also direct \textit{Interstellar}?''

\emph{Implementation (already completed).} Extraction had produced the bridging triples (\textit{Inception}, \texttt{directed\_by}, \textit{Christopher Nolan}) and (\textit{Interstellar}, \texttt{directed\_by}, \textit{Chris Nolan}), both with high confidence ($\tcs\approx0.94$), and $\scr=0.91$ over the source passages, so extraction was not the weak link.

\emph{Validation (already completed).} On the assembled graph $\dnr=0.18$ was mildly elevated because the surface form \textit{Chris Nolan} had resolved to its own near-isolated node.

\emph{Usage (at inference).} For this query $\qgr=1.0$ (both films are nodes) and $\erc=1.0$ (both films, the query entities, are retrieved), yet $\rrs=0$: the two films sit in the retrieved subgraph with no path between them, because their shared director is split across two unmerged nodes.

\emph{Diagnosis and remediation.} Reading the chain (\cref{fig:chain}), the first broken link is \rrs, a \emph{connectivity gap}; the indicated remediation is a graph-and-schema lever, improved entity resolution to merge \textit{Chris Nolan} into \textit{Christopher Nolan}, which restores the two-hop path.
\end{minipage}}
\caption{A single query threaded through the three \tri phases. Each phase
reports one headline metric; the earlier phases ran offline, while the chain is
read at inference. The first broken link (\cref{fig:chain}) localizes the failure
to a single condition, here a connectivity gap, which then indicates the stage
lever that can fix it. Values are illustrative (MetaQA / WikiMovies).}
\label{fig:worked}
\end{figure}

\paragraph{Takeaways.} Four practical points summarize the framework.
First, localize a failure to a stage and act on it with that stage's levers, rather than only observing it at the answer.
Second, do not wait for gold labels: the core signals are computable in deployment on the artifacts a pipeline already has.
Third, measure the retrieved subgraph, not only the generated answer, so a retrieval failure cannot hide behind a fluent response.
Fourth, triage before generation, since the first three chain links are available before the LLM runs.
If only a handful of signals can be monitored, we suggest \tcs (extraction confidence), \dnr (graph connectivity), and the chain heads \qgr, \erc, and \rrs (scope, retrieval, connectivity).

\section{Trust Metrics}
\label{sec:metrics}

This section formally defines the \tri metric suite, summarized by phase in \cref{tab:metrics}.
We first fix a shared notation (\cref{tab:notation}), then introduce the two axes that govern which metrics are computable in a given setting, and finally present the metrics phase by phase, each followed by a short note on the remediation it triggers.
Metrics marked [TRIAGE] are defined or operationalized in this work for the TRIAGE framework; several instantiate standard measurement ideas, such as coverage, consistency, schema violation, precision/recall, or path connectivity, in a stage-localized graph-RAG setting; the contribution is their joint organization into a deployable diagnostic chain and remediation map. All others are imported using their original definitions and cite their source, whether the \tkg methodology~\cite{TKG-Amdouni2026} or an external graph-RAG benchmark.

\begin{table}[t]
\centering
\caption{Shared notation used throughout \cref{sec:metrics}.}
\label{tab:notation}
\small
\begin{tabular}{ll}
\toprule
\textbf{Symbol} & \textbf{Meaning} \\
\midrule
$\hat{t}=(s,p,o)$ & an \emph{extracted} triple (hat denotes extracted, vs.\ gold $t$) \\
$E=\{\hat{t}_j\}$ & set of extracted triples for a document or corpus \\
$G=\{t_i\}$ & set of gold triples (when available) \\
$\KG$ & triple set of the assembled TRIAGE-KG (assessed graph) \\
$\refKG$ & triple set of the reference KG (benchmark setting) \\
$V$ & vertex (entity) set of the assembled KG $\KG$ \\
$\ont,\ \ont_{\mathrm{rel}}$ & reference ontology and its query-relevant subset \\
$f(\cdot)$ & text-embedding function \\
$\sgrel(x,y)=\cos(f(x),f(y))$ & component-level similarity built from $f$ \\
$\mathcal{P}=\{p_1,\dots,p_n\}$ & set of input passages \\
$\KG_S,\ V_S$ & retrieved subgraph: triples $\KG_S\subseteq\KG$, vertices $V_S\subseteq V$ \\
\(M_q\) & surface mentions extracted from query \(q\) \\
\(\gamma_q\) & grounding function \(M_q \to V \cup \{\bot\}\) \\
\(V_q^g\) & grounded query-node set \(\{\gamma_q(m): m\in M_q,\ \gamma_q(m)\neq\bot\}\subseteq V\) \\
\(M_a\) & surface mentions extracted from generated answer \(A\) \\
\(\gamma_a\) & grounding function \(M_a \to V \cup \{\bot\}\) \\
\(V_a^g\) & grounded answer-node set
\(\{\gamma_a(m): m\in M_a,\ \gamma_a(m)\neq\bot\}\subseteq V\) \\
$\deg(v)$ & degree of node $v$ \\
$\tau,\ \theta$ & confidence / decision thresholds \\

\bottomrule
\end{tabular}
\end{table}

The suite is generated by crossing three trust questions, namely whether each \emph{triple} is sound, whether the assembled \emph{graph} is sound, and whether \emph{usage} succeeds, with the two computability axes introduced next.
A metric earns its place by occupying a cell with a distinct failure mode and a distinct remediation, which is also why we report no single composite score: graph quality for retrieval is multi-objective, with trade-offs such as connectivity against parsimony that a scalar would hide.
Most of the metrics are individually simple and several are adapted from prior work; what is new is their organization into a stage-localized instrument whose core signals need no gold annotations at deployment.

Two orthogonal axes determine which metrics can be computed in a given setting, and we use them to organize the suite.

\textbf{Reference requirement.} The first axis is the \emph{external reference} a metric needs, which is what governs its availability in a given setting.
Most metrics need \emph{none}: they are gold-free and schema-free, computed on any KG from the artifacts the pipeline already produces (\tcs, \ec, \csrlink, \dnr at construction; \qgr, \erc, \rrs at usage).
A second group needs a reference \emph{ontology} (marked $\S$: \oc, \scomp, \cvr, \LCm, and ontology-constrained ranking $\mathrm{MRR}_{\mathrm{type}}$); since an ontology can be supplied at deployment, these stay live wherever a schema is on hand and become undefined otherwise, the measurement-side counterpart of the two canonicalization regimes of \cref{sec:layer1}.
A third group needs a \emph{gold standard} (marked $\dagger$, e.g.\ \tfi, \CORm, \CMPm, \gpc, and the imported outcome metrics \ctxcov, \rfr, \hhr); gold is never available at inference, so these are computable only offline and sit below the dashed line in each phase of \cref{tab:metrics}, acting as yardsticks for the gold-free metrics (for example \gpc against \erc and \rrs) rather than as deployment signals.

\textbf{Visibility: white-box vs black-box.} \tri is a white-box instrument: each phase's metrics require observing that phase's internal artifact, and with only black-box (input and output) access they are unavailable (\cref{tab:visibility}).
At \tkgterm{KG Implementation}, white-box access means observing the extraction process, its source text, token log-probabilities, and run variance, which enables \tcs, \scr, \ec, \xsc, \ear, \csrlink, and the fidelity metrics; a black-box graph supplied by a third party exposes none of these, and link prediction (MRR, $\mathrm{MRR}_{\mathrm{type}}$, ECE) becomes the completeness probe instead (\cref{sec:metrics-kgc}).
At \tkgterm{KG Usage}, white-box access means observing the retrieved subgraph, the mapped query entities, and the connecting paths, which is what makes \qgr, \erc, \rrs, and the answer-side metrics computable; with only the query and the final answer, one is back to answer correctness (Hits@1/F1) and LLM-as-judge answer-level dimensions.
\tkgterm{KG Validation} always observes the assembled graph, so its metrics do not depend on visibility.
\Cref{tab:metrics} marks ontology-dependent metrics with $\S$ and sets off the gold-dependent ones ($\dagger$) below a dashed line within each phase; \cref{tab:visibility} summarizes the per-phase visibility requirements, and \cref{tab:corederived} separates primary signals from derived ones.
A first reading can follow the \emph{primary} signals that form the spine of the suite (\cref{tab:corederived}) and treat the derived, aggregate, and imported metrics as optional diagnostic detail. 
\begin{table}[tbp]
\centering
\caption{Overview of the \tri confidence metric suite by \tkg phase.
  \textbf{[TRIAGE]}: defined or operationalized here for the TRIAGE framework; imported metrics cite their source.
  $(\S)$ requires an ontology; $(\dagger)$ requires a gold reference.
  Within each phase, the metrics above the dashed line are gold-free and form the
  deployable instrument; those below ($\dagger$) need a gold reference and are
  used only offline, to calibrate and validate the gold-free metrics.
  The link-prediction rows (MRR, Hits@K, $\mathrm{MRR}_{\mathrm{type}}$, ECE) are a
  black-box completeness probe requiring held-out triples, not gold-free
  deployment signals.}
\label{tab:metrics}
\setlength{\tabcolsep}{5pt}
\renewcommand{\arraystretch}{1.05}
\footnotesize
\begin{tabular}{llll}
\toprule
\textbf{Phase} & \textbf{Metric} & \textbf{Symbol} & \textbf{Origin} \\
\midrule
\multirow{16}{*}{KG Implementation}
  & Triple Confidence Score & \tcs & [TRIAGE] \\
  & Source Coverage Ratio & \scr & [TRIAGE] \\
  & Extraction Consistency & \ec & [TRIAGE] \\
  & Cross-Source Corroboration & \xsc & [TRIAGE] \\
  & Entailment Ratio & $\mathrm{EntRatio}$ & \cite{honovich2022true} \\
  & Schema Compliance Rate$^{\S}$ & \scomp & [TRIAGE] \\
  & Constraint Violation Rate$^{\S}$ & \cvr & [TRIAGE] \\
  & Canonicalization Success Rate & \csrlink & [TRIAGE] \\
  & Evidence Attribution Rate & \ear & [TRIAGE] \\
  & Aggregated Trustworthiness Index & \triti & [TRIAGE] \\
  & Mean Rank / MRR / Hits@K & $\mathrm{MRR}$ & \cite{bordes2013transe} \\
  & Ontology-constrained ranking$^{\S}$ & $\mathrm{MRR}_{\mathrm{type}}$ & [TRIAGE] \\
  & Calibration of completion conf. & $\mathrm{ECE}$ & \cite{guo2017calibration} \\
\cdashline{2-4}
  & Embedding Similarity$^{\dagger}$ & $\cos(\hat t,t)$ & \cite{reimers2019sentencebert} \\
  & Soft-F1$^{\dagger}$ & $\mathrm{Soft\text{-}F1}$ & \cite{bhardwaj2019carb} \\
  & Triple Fidelity Index$^{\dagger}$ & \tfi & [TRIAGE] \\
\midrule
\multirow{6}{*}{KG Validation}
  & Logical Consistency$^{\S}$ & \LCm & \cite{TKG-Amdouni2026} \\
  & Ontology Coverage$^{\S}$ & \oc & [TRIAGE] \\
  & Dead Node Ratio & \dnr & [TRIAGE] \\
  & Semantic Redundancy Rate & \srr & [TRIAGE] \\
\cdashline{2-4}
  & Correctness$^{\dagger}$ & \CORm & \cite{TKG-Amdouni2026} \\
  & Completeness$^{\dagger}$ & \CMPm & \cite{TKG-Amdouni2026} \\
\midrule
\multirow{16}{*}{KG Usage}
  & Query Grounding Rate & \qgr & [TRIAGE] \\
  & Entity Retrieval Coverage & \erc & [TRIAGE] \\
  & Entity Retrieval Precision & \erp & [TRIAGE] \\
  & Reasoning Readiness Score & \rrs & [TRIAGE] \\
  & Retrieval Path Cost & \rpc & [TRIAGE] \\
  & Answer Grounding Rate & \agr & [TRIAGE] \\
  & Answer Utilization Rate & \aur & [TRIAGE] \\
  & Answer Reasoning Faithfulness & \arf & [TRIAGE] \\
  & Reasoning Hop Depth & \rhdm & [TRIAGE] \\
  & KG Net Impact & \kgimpact & \cite{eval-finhallubench2026} \\
\cdashline{2-4}
  & Gold Path Coverage$^{\dagger}$ & \gpc & [TRIAGE] \\
  & Retrieval Path Precision/Recall$^{\dagger}$ & $\mathrm{RPP/RPR}$ & [TRIAGE] \\
  & Answer Correctness$^{\dagger}$ & $\mathrm{Hits@1/F1}$ & \cite{yih2016webqsp} \\
  & Context Coverage$^{\dagger}$ & \ctxcov & \cite{eval-reasoningbottleneck2026} \\
  & Reasoning Failure Rate$^{\dagger}$ & \rfr & \cite{eval-reasoningbottleneck2026} \\
  & Hard Hits Rate$^{\dagger}$ & \hhr & \cite{eval-whatbreakskgrag2025} \\
\bottomrule
\end{tabular}
\end{table}

\begin{table}[tbp]
\centering
\caption{\tri is a \emph{white-box} instrument: each phase's metrics require
  observing that phase's internal artifact. With only \emph{black-box} access
  (the phase's inputs and outputs) those metrics are unavailable, and one falls
  back to the right-hand column. Visibility is independent per phase, and is
  separate from ontology-dependence, marked $\S$ in \cref{tab:metrics}.}
\label{tab:visibility}
\small
\setlength{\tabcolsep}{4pt}
\renewcommand{\arraystretch}{1.35}
\begin{tabular}{@{}l p{0.40\linewidth} p{0.34\linewidth}@{}}
\toprule
\textbf{Phase} & \textbf{White-box artifact (metrics it enables)} & \textbf{Black-box fallback} \\
\midrule
Implementation & extraction process: source text, token log-probabilities, run variance ($\Rightarrow$ \tcs, \scr, \ec, \xsc, \ear, fidelity) & only the output triples: link prediction (MRR, Hits@K, $\mathrm{MRR}_{\mathrm{type}}$, ECE) \\
Validation & the assembled graph ($\Rightarrow$ \dnr, \srr, \LCm, \oc) & the graph is the artifact under review, so it is observed by construction \\
Usage & the retrieved subgraph, mapped query entities, and connecting paths ($\Rightarrow$ \qgr, \erc, \erp, \rrs, \agr, \aur, \arf, \rhdm, \rpc) & only the query and the answer: answer correctness (Hits@1/F1) and LLM-as-judge answer-level dimensions \\
\bottomrule
\end{tabular}
\end{table}

\begin{table}[tbp]
\centering
\caption{Dependency/redundancy analysis. \emph{Primary} metrics are
  independent signals worth triggering remediation on; \emph{derived} metrics
  are definitional complements, aggregates, or expected correlates of a primary
  metric, and serve as diagnostic detail. No claim of statistical independence
  is made; relationships are definitional or hypothesized
  (\cref{sec:metrics-depend}).}
\label{tab:corederived}
\small
\begin{tabular}{lll}
\toprule
\textbf{Metric} & \textbf{Role} & \textbf{Relationship} \\
\midrule
\tcs & primary & extraction confidence \\
\scr & primary & source coverage \\
\ec & primary & run-to-run stability \\
\xsc & primary & cross-source agreement \\
\scomp & primary & schema validity \\
\cvr & derived & per-constraint refinement of \scomp \\
\csrlink & primary & grounding success \\
\triti, \tfi & derived & weighted aggregates of the above \\
\midrule
\LCm & primary & graph-level consistency \\
\oc & primary & schema coverage \\
\dnr & primary & structural inertia \\
\srr & primary & structural redundancy \\
\CMPm & derived & gold-anchored counterpart of \oc \\
\midrule
\qgr & primary & query in-scope for KG \\
\erc & primary & retrieval recall of grounded query nodes \\
\erp & derived & precision complement of \erc \\
\rrs & primary & subgraph connectivity \\
\agr & primary & answer grounding \\
\aur & derived & recall complement of \agr \\
\arf & primary & relational faithfulness \\
\rpc & primary & retrieval cost \\
\rhdm & derived & path-length distribution behind \rrs \\
\bottomrule
\end{tabular}
\end{table} 
\subsection{KG Implementation Metrics}
\label{sec:metrics-layer1}

The \tkgterm{KG Implementation} phase produces a set of extracted triples $E=\{\hat{t}_j\}$, each $\hat{t}=(s,p,o)$ a candidate fact from the source documents.
Evaluating them requires three complementary perspectives.
\emph{Triple fidelity} metrics assess correctness relative to gold triples or source evidence (and so require $G$ or source text $P$).
\emph{Trustworthiness} metrics are intrinsic signals computable without gold annotations.
\emph{KG completion} metrics characterize the structural quality of the graph and contextualize why trustworthiness signals are needed alongside accuracy.

\emph{Core question:} is each extracted triple trustworthy?
The signal to watch first is the Triple Confidence Score (\tcs); the other Implementation metrics refine and complement it.

\subsubsection{Triple Fidelity Metrics}
\label{sec:metrics-fidelity}

Triple fidelity metrics measure how well extracted triples match gold knowledge or source evidence; they are benchmark-setting metrics, applicable when a reference set $G$ or source text $P$ is available.
The encoder $f$ is a free parameter: sentence-transformer encoders are a fast, inexpensive default well suited to short triple strings, while LLM-based embedders capture relational nuance at higher cost and latency.
The metric is only as reliable as the encoder's ability to place semantically equivalent triples near one another.

\paragraph{Embedding similarity.}
Given the embedding function $f$, the cosine similarity between an extracted
triple $\hat{t}$ and a gold triple $t$
is~\cite{reimers2019sentencebert,salton1983ir}
\begin{equation}
  \cos(\hat{t}, t) = \frac{f(\hat{t}) \cdot f(t)}
    {\|f(\hat{t})\|\,\|f(t)\|}.
\end{equation}
Each triple is linearized into a string (e.g., \texttt{s p o}) before
embedding, capturing semantic relatedness even when triples differ lexically.

\paragraph{Soft matching and Soft-F1.}
Exact matching is too strict for LLM outputs.
Using the shared component similarity $\sgrel(x,y)=\cos(f(x),f(y))$ from
\cref{tab:notation}, component-wise triple similarity
is~\cite{bhardwaj2019carb,stanovsky2016oiebenchmark}
\begin{equation}
  \mathrm{Score}(\hat{t}, t) = \tfrac{1}{3}\big(
    \sgrel(s_{\hat{t}}, s_t) +
    \sgrel(p_{\hat{t}}, p_t) +
    \sgrel(o_{\hat{t}}, o_t)\big).
\end{equation}
Set-level Soft-Precision and Soft-Recall are computed via maximum-weight
bipartite matching $M$ between $E$ and $G$~\cite{kuhn1955hungarian}:
\begin{equation}
  \mathrm{Soft\text{-}P} =
    \frac{1}{|E|}\!\!\sum_{(\hat{t},t) \in M}\!\! \mathrm{Score}(\hat{t}, t),
  \qquad
  \mathrm{Soft\text{-}R} =
    \frac{1}{|G|}\!\!\sum_{(\hat{t},t) \in M}\!\! \mathrm{Score}(\hat{t}, t),
\end{equation}
\begin{equation}
  \mathrm{Soft\text{-}F1} =
    \frac{2\, \mathrm{Soft\text{-}P}\cdot \mathrm{Soft\text{-}R}}
    {\mathrm{Soft\text{-}P} + \mathrm{Soft\text{-}R}}.
\end{equation}

\paragraph{Entailment-based verification.}
To measure grounding in source evidence, triple validation is reframed as
Natural Language Inference (NLI).
NLI classifies a (premise, hypothesis) pair into one of three labels:
\emph{entailment} (the premise supports the hypothesis), \emph{contradiction}
(the premise refutes it), or \emph{neutral} (neither).
Here the premise is the \emph{source passage} $p(\hat{t})$ from which $\hat{t}$
was extracted (its recorded provenance from \cref{sec:layer1}, not the whole corpus), and the hypothesis $H(\hat{t})$ is a natural-language verbalization of
the triple (e.g., $(s,p,o)\mapsto$ ``$s$ $p$ $o$'').
Writing $P_{\mathrm{NLI}}(\text{ent}\mid p(\hat{t}),H)$ for the probability mass
an NLI model assigns to the entailment label, the corpus-level entailment ratio
is~\cite{bowman2015snli,williams2018multinli,honovich2022true}
\begin{equation}
  \mathrm{EntRatio}(E) =
    \frac{1}{|E|} \sum_{\hat{t} \in E}
    P_{\mathrm{NLI}}\!\big(\text{ent} \mid p(\hat{t}), H(\hat{t})\big).
\end{equation}
Triples with low entailment and high contradiction probability are flagged
as unsupported, a grounding signal complementary to Soft-F1.

\paragraph{Triple Fidelity Index ($\tfi$) \textbf{[TRIAGE]}, benchmark-only.}
To summarize the three fidelity signals into a single $[0,1]$ score,
paralleling the trustworthiness aggregate \triti below:
\begin{equation}
  \tfi(E) = \lambda_1\, \overline{\cos} + \lambda_2\, \mathrm{Soft\text{-}F1}
            + \lambda_3\, \mathrm{EntRatio},
  \quad \textstyle\sum_i \lambda_i = 1,
\end{equation}
where $\overline{\cos}$ is the mean best-match cosine over $E$ and the
weights $\lambda_i$ reflect the relative importance of semantic match,
set-level agreement, and source grounding.
Because two of its three components ($\overline{\cos}$ and Soft-F1) require the
gold set $G$, \tfi is a benchmark-setting aggregate and is \emph{not} computable
in gold-free deployment, even though its third component, EntRatio, needs only
the source passages $P$.

\subsubsection{TRIAGE Trustworthiness Metrics}
\label{sec:metrics-trust}

Unlike fidelity metrics, trustworthiness metrics are \emph{intrinsic}: they require neither gold triples nor source alignment, and so are operational in any deployment.
They capture extraction confidence (\tcs), source coverage (\scr), run-to-run stability (\ec), cross-source agreement (\xsc), structural validity (\scomp, \cvr), and grounding to canonical identifiers (\csrlink).

\paragraph{Triple Confidence Score ($\tcs$) \textbf{[TRIAGE]}.}
LLM token probabilities provide an intrinsic proxy for extraction
confidence~\cite{bengio2003neuralLM,kadavath2022know,jiang2021when}.
Treating the emitted triple as its token sequence $w_1,\dots,w_{|\hat t|}$,
where $w_k$ is the $k$-th token and $w_{<k}$ its predecessors, and writing
$P_{\mathrm{LLM}}(w_k\mid w_{<k})$ for the model's next-token probability, we
define \tcs as the \emph{geometric mean} token probability, equivalently the
exponentiated mean token log-probability:
\begin{equation}
  \tcs(\hat{t}) =
    \exp\!\Big(\frac{1}{|\hat{t}|} \sum_{k=1}^{|\hat{t}|}
    \log P_{\mathrm{LLM}}(w_k \mid w_{<k})\Big)
    \;=\;\Big(\textstyle\prod_{k=1}^{|\hat{t}|}
      P_{\mathrm{LLM}}(w_k\mid w_{<k})\Big)^{1/|\hat{t}|}
    \;\in(0,1].
\end{equation}
Under this convention \emph{higher \tcs is better} (the model was more
confident), the score is length-normalized and bounded, and a triple is flagged
when $\tcs(\hat{t})<\tau$ for a threshold $\tau$.
Being already in $(0,1]$, \tcs enters the aggregate index below directly, with no
rescaling.
\tcs requires access to token log-probabilities from the extracting LLM (white-box setting); it is undefined for a black-box graph.

\paragraph{Source Coverage Ratio ($\scr$) \textbf{[TRIAGE]}.}
Correctness does not imply completeness.
\scr measures how broadly the input yields at least one confident triple.
With passages $\mathcal{P}=\{p_1,\dots,p_n\}$ and threshold $\tau$:
\begin{equation}
  \scr =
    \frac{\big|\{p_i \in \mathcal{P} :
      \exists\, \hat{t} \in E(p_i)\ \text{s.t.}\ \tcs(\hat{t}) \ge \tau\}\big|}
    {|\mathcal{P}|}\;\in[0,1].
\end{equation}
Low \scr indicates that large portions of the input fail to yield confident
triples, mapping to the \tkgterm{Knowledge Elicitation} coverage notion
of~\cite{TKG-Amdouni2026}.

\paragraph{Extraction Consistency ($\ec$) \textbf{[TRIAGE]}.}
LLM extraction is sensitive to stochastic decoding~\cite{wang2023selfconsistency}.
Given $k$ independent extractions $E^{(1)},\dots,E^{(k)}$ of the same input, \ec
is the average pairwise Jaccard similarity~\cite{jaccard1901}:
\begin{equation}
  \ec =
    \frac{2}{k(k-1)} \sum_{i < j}
    \frac{|E^{(i)} \cap E^{(j)}|}{|E^{(i)} \cup E^{(j)}|}\;\in[0,1].
\end{equation}
A semantic variant replaces set overlap with Soft-F1 matching to absorb
paraphrastic variation across runs.

\paragraph{Cross-Source Corroboration ($\xsc$) \textbf{[TRIAGE]}.}
Whereas \ec measures agreement across \emph{runs} of the same input, \xsc
measures agreement across \emph{independent sources}: a triple extracted
independently from two or more documents is far more likely to be true.
Let $\mathrm{src}(\hat{t})$ be the set of distinct source documents from which a
(canonicalized) triple $\hat{t}$ is extracted.
Then
\begin{equation}
  \xsc(E) =
    \frac{|\{\hat{t}\in E : |\mathrm{src}(\hat{t})| \ge 2\}|}{|E|}\;\in[0,1].
\end{equation}
A higher \xsc indicates a graph whose facts are corroborated across the
corpus rather than resting on single, possibly idiosyncratic, mentions.
\xsc requires multi-document provenance and therefore applies in the
white-box setting.

\paragraph{Schema Compliance Rate ($\scomp$) \textbf{[TRIAGE]}.}
Schema compliance measures conformance to the ontology's relation vocabulary
and type constraints~\cite{nickel2015review,paulheim2017kgquality}.
With predicate set $\mathcal{R}$, domain/range constraints
$\mathrm{dom}(p),\mathrm{rng}(p)$, and inferred type $\mathrm{type}(x)$:
\begin{equation}
  \label{eq:sc-valid}
  \mathrm{valid}_{\ont}(\hat{t}) =
    \mathbb{I}[p_{\hat{t}}\!\in\!\mathcal{R}]\cdot
    \mathbb{I}[\mathrm{type}(s_{\hat{t}})\!\in\!\mathrm{dom}(p_{\hat{t}})]\cdot
    \mathbb{I}[\mathrm{type}(o_{\hat{t}})\!\in\!\mathrm{rng}(p_{\hat{t}})],
\end{equation}
\begin{equation}
  \scomp(E) = \frac{1}{|E|}\sum_{\hat{t}\in E}\mathrm{valid}_{\ont}(\hat{t})
  \;\in[0,1].
\end{equation}
\scomp requires an ontology and is undefined without one.

\paragraph{Constraint Violation Rate ($\cvr$) \textbf{[TRIAGE]}.}
Where \scomp is a per-triple pass/fail signal, \cvr quantifies how
frequently each constraint type is violated.
The constraint set is exactly the three checks composing \scomp in \cref{eq:sc-valid}, $\mathcal{C}=\{\texttt{pred},\texttt{dom},\texttt{rng}\}$ (predicate membership, domain, and range), 
with \(\mathrm{viol}_c(\hat{t}) = 1 - I_c(\hat{t})\), where \(I_c(\hat{t})\) denotes the \(c\)-th indicator composing \(\mathrm{valid}_{O}(\hat{t})\),
so that \cvr decomposes \scomp by constraint type rather than introducing new
checks:
\begin{equation}
  \cvr(E) = \frac{1}{|E|}\sum_{\hat{t}\in E}
    \Big(\frac{1}{|\mathcal{C}|}\sum_{c\in\mathcal{C}}\mathrm{viol}_c(\hat{t})\Big)
  \;\in[0,1].
\end{equation}
\cvr can be reported per constraint type (e.g.,
$\cvr_{\texttt{dom}},\cvr_{\texttt{rng}}$) to localize systematic schema drift;
it too requires an ontology.

\paragraph{Canonicalization Success Rate ($\csrlink$) \textbf{[TRIAGE]}.}
A triple may be correct at the string level yet unusable if its entities
cannot be grounded to canonical
identifiers~\cite{shen2015entitylinking,ji2014tackbpedl}.
Let $C(\cdot)$ map surface forms to canonical identifiers, returning $\bot$ when
no link is found:
\begin{equation}
  \csrlink(E) = \frac{1}{|E|}\sum_{\hat{t}\in E}
    \mathbb{I}[C(s_{\hat{t}})\neq\bot]\cdot
    \mathbb{I}[C(o_{\hat{t}})\neq\bot]\;\in[0,1].
\end{equation}
The meaning of ``canonical'' depends on the regime of \cref{sec:layer1}.
In the \emph{ontology-grounded} regime, $C$ maps to schema or KB identifiers and
\csrlink measures external grounding success.
In the \emph{resolution-based} regime (no ontology), $C$ maps to the internally
induced canonical set produced by entity resolution, and \csrlink measures
internal consistency rather than external grounding.
This is the point at which the no-ontology consequence becomes concrete: when
\oc, \scomp, \LCm, and \CMPm are undefined for lack of a schema, the
trustworthiness load shifts onto \csrlink, \dnr, \ec, and \tcs.

\paragraph{Evidence Attribution Rate ($\ear$) \textbf{[TRIAGE]}, optional.}
For auditability, \ear measures how often a provenance evidence span is
available for a triple, enabling the verification motivated in
\cref{sec:related}.
With $\mathrm{span}(\hat{t})$ the (possibly empty) evidence set:
\begin{equation}
  \ear(E) = \frac{1}{|E|}\sum_{\hat{t}\in E}
    \mathbb{I}[\mathrm{span}(\hat{t})\neq\emptyset]\;\in[0,1].
\end{equation}

\paragraph{Aggregated Trustworthiness Index ($\triti$) \textbf{[TRIAGE]}, optional.}
To summarize complementary trustworthiness signals into one indicator
(all terms already lie in $[0,1]$ and increase with quality):
\begin{equation}
\begin{aligned}
\triti(E)
&= \omega_1 TCS + \omega_2 SCR + \omega_3 EC \\
&\quad + \omega_4 CSC + \omega_5 SC + \omega_6 CSR_{\mathrm{link}},
&\qquad \sum_i \omega_i = 1 .
\end{aligned}
\end{equation}
where weights $\omega_i$ reflect application risk (e.g., higher $\omega_5$ in
ontology-critical domains).
Because \tcs is now bounded in $(0,1]$, it enters the sum directly with no
rescaling, so \triti is comparable across datasets and runs.
Each component remains independently interpretable.
\triti is gold-free but not schema-free: its \scomp term is undefined without an
ontology.
When no ontology is available, \scomp is dropped and the remaining weights are
renormalized to sum to one, so \triti stays computable and comparable on
the schema-free subset.

\paragraph{Remediation note (Implementation).} Low Implementation metrics call for \emph{extraction levers}, which leave graph structure and retrieval untouched: refine the extraction prompt, lower decoding temperature, add or rebalance few-shot demonstrations, or re-chunk the input.
For example, low \scr or \xsc points to under-extraction or single-source fragility (acquire or re-process documents); low \ec points to decoding instability (reduce temperature, self-consistency voting); low \scomp/\cvr points to schema drift (tighten the prompt or predicate inventory).
These are the extraction levers of the remediation map (\cref{sec:remediation}).

\subsubsection{KG Completion (KGC) Metrics}
\label{sec:metrics-kgc}

Knowledge Graph Completion (KGC), or link prediction, evaluates a model's
ability to predict missing \emph{entities} in an existing graph: given
$(s,p,?)$ or $(?,p,o)$, the relation fixed, it ranks candidate entities so
the correct one appears near the top~\cite{bordes2013transe}.
(The related task of relation prediction $(s,?,o)$ exists but is not the setup
these rank-based metrics target.)
As discussed above, KGC is primarily relevant in the
\emph{black-box} regime: when a graph arrives without source
text, log-probabilities, or extraction provenance, the extraction-oriented
trustworthiness metrics are undefined, and the meaningful question shifts from
``was this graph extracted well?'' to ``is this graph complete?'', which is precisely what link prediction probes.
Reporting is in the \emph{filtered} setting~\cite{bordes2013transe}:
\begin{itemize}
  \item \textbf{Mean Rank}: $\mathrm{MR}=\frac{1}{|Q|}\sum_{q}\mathrm{rank}_q$;
  \item \textbf{MRR}~\cite{trouillon2016complex}:
    $\mathrm{MRR}=\frac{1}{|Q|}\sum_{q}\frac{1}{\mathrm{rank}_q}$;
  \item \textbf{Hits@K}~\cite{sun2019rotate}:
    $\mathrm{Hits@K}=\frac{1}{|Q|}\sum_{q}\mathbb{I}[\mathrm{rank}_q\le K]$,
\end{itemize}
with $\mathrm{rank}_q$ the rank of the correct entity and $K\in\{1,3,10\}$.

\paragraph{Ontology-constrained ranking ($\mathrm{MRR}_{\mathrm{type}}$) \textbf{[TRIAGE]}.}
Standard KGC ranks the correct entity against \emph{all} candidates, including
type-incompatible ones that the schema already rules out.
When an ontology is available, \tri restricts the candidate set to the
type-valid entities for the relation, so that completion is judged on
\emph{semantically plausible} candidates rather than trivially excluded ones.
Let $\mathcal{E}_{\mathrm{type}}(p)$ be the set of entities compatible with the
domain or range of relation $p$ under the ontology, and let
$\mathrm{rank}^{\mathrm{type}}_q$ be the rank of the correct entity within
$\mathcal{E}_{\mathrm{type}}(p)$:
\begin{equation}
  \mathrm{MRR}_{\mathrm{type}} =
    \frac{1}{|Q|}\sum_{q\in Q}\frac{1}{\mathrm{rank}^{\mathrm{type}}_q}.
\end{equation}
$\mathrm{MRR}_{\mathrm{type}}$ ties link prediction to the same schema axis as
\scomp and \cvr: it rewards a model for placing the correct entity ahead of
other \emph{type-valid} alternatives, not merely ahead of entities the ontology
would reject anyway. It requires an ontology and is undefined without one.

\paragraph{Calibration of completion confidence ($\mathrm{ECE}$), adopted.}
A black-box graph is sometimes accompanied by per-edge confidence
scores, or a completion model produces them; their usefulness depends on
whether they reflect true correctness likelihood.
We adopt the Expected Calibration Error~\cite{guo2017calibration} to assess
this, binning predictions into $B_1,\dots,B_M$ by confidence and comparing
per-bin accuracy with confidence:
\begin{equation}
  \mathrm{ECE} =
    \sum_{m=1}^{M}\frac{|B_m|}{n}\,
    \big|\mathrm{acc}(B_m)-\mathrm{conf}(B_m)\big|.
\end{equation}
ECE is a standard metric rather than a contribution of this work; what \tri
adds is its \emph{role} in the framework: a low ECE licenses using a graph's
confidence scores as a trust signal at \tkgterm{KG Validation}, while a high ECE
warns that those scores cannot be trusted and the schema-free structural metrics
(\dnr, \srr) should carry the assessment instead.
This mirrors, on the black-box side, the calibration concern that affects \tcs
on the white-box side (\cref{sec:metrics-trust}).

\paragraph{Limitations.} Three limitations motivate caution: rank metrics depend on candidate-set size, hampering cross-dataset comparison~\cite{hoyt2022rankmetrics,ruffinelli2020you}; closed-world evaluation treats unobserved triples as negatives~\cite{bordes2013transe}; and benchmark artifacts such as inverse-relation leakage inflate scores~\cite{dettmers2018conve,toutanova2015observed}.
\tri therefore complements KGC reporting with the intrinsic trustworthiness metrics above rather than relying on rank metrics alone. 
\subsection{KG Validation Metrics}
\label{sec:metrics-layer2}

The \tkgterm{KG Validation} phase assesses the assembled KG before use.
A key shift from \cref{sec:metrics-layer1} is one of \emph{level}: where Implementation metrics are computed per triple, Validation metrics are computed over the whole graph $\KG$.
Logical consistency is the clearest example: here it is a \emph{global} property of $\KG$ under ontology axioms, detected by a reasoner over the entire graph, not a per-triple schema check.
Validation metrics fall into two groups: \emph{ground-truth-free} metrics, computable in any deployment, and \emph{benchmark} metrics requiring a reference KG $\refKG$.

\emph{Core question:} does the assembled graph hold together well enough to query?
The schema-free signal to watch first is the Dead Node Ratio (\dnr), with Logical Consistency (\LCm) and Ontology Coverage (\oc) adding schema-based checks when an ontology is available.

\medskip\noindent\textbf{Ground-truth-free metrics.}

\paragraph{Logical Consistency (\LCm, $\mlc$).}
Absence of internal contradictions in $\KG$, detected by OWL reasoning
against the ontology~\cite{TKG-Amdouni2026}.
Let $\mathrm{Conf}(\KG)\subseteq\KG$ be the set of triples that participate in at
least one violated ontology axiom; defining the numerator as this triple count
(rather than the number of violation \emph{instances}, which can exceed $|\KG|$)
keeps the score in $[0,1]$:
\begin{equation}
  \mlc = 1 - \frac{|\mathrm{Conf}(\KG)|}{|\KG|}\;\in[0,1],
\end{equation}
where a triple is in $\mathrm{Conf}(\KG)$ if it occurs in some triple set
violating an ontology axiom (disjointness, functionality, cardinality), as
found by a reasoner such as HermiT over the full graph.
\LCm requires an ontology.

\paragraph{Ontology Coverage ($\oc$) \textbf{[TRIAGE]}.}
\oc measures what fraction of the query-relevant ontology is instantiated in
$\KG$:
\begin{equation}
  \oc =
    \frac{|\{c \in \ont_{\mathrm{rel}} :
      \exists\, t \in \KG\ \text{instantiating}\ c\}|}
    {|\ont_{\mathrm{rel}}|}\;\in[0,1].
\end{equation}
\oc is the structural-completeness signal of this phase: not whether triples
are correct (that is \CORm) but whether the graph covers the domain as
defined by the ontology.
Low \oc on a class suggests either missing source content or extraction failure
for that class, both actionable.

\paragraph{Dead Node Ratio ($\dnr$) \textbf{[TRIAGE]}.}
A \emph{dead node} has degree $\le 1$: isolated (degree 0) or a dangling
leaf (degree 1).
Such nodes cannot serve as intermediate bridges in multi-hop reasoning: a degree-1 node may still be a valid path endpoint (e.g.\ an answer entity), but
it cannot connect two other entities along a reasoning chain.
With node set $V$ of $\KG$:
\begin{equation}
  \dnr = \frac{|\{v\in V : \deg(v)\le 1\}|}{|V|}\;\in[0,1].
\end{equation}
High \dnr signals disconnected entities: surface-form variants, over-segmented mentions, or hallucinated entities with no relational context.
It complements \LCm (contradictions) and \oc (missing concepts) by detecting
\emph{structurally inert} content, and is computable without an ontology.

\paragraph{Semantic Redundancy Rate ($\srr$) \textbf{[TRIAGE]}.}
Where \dnr detects too little connectivity, \srr detects too much
duplication: triples that are semantically equivalent or subsumed by others,
inflating the graph without adding information.
Using the component similarity $\sgrel$ and a threshold $\delta$, call two
triples \emph{redundant} if all three components match above $\delta$; let
$\mathrm{Red}(\KG)$ be the set of triples that duplicate or are subsumed by
another:
\begin{equation}
  \srr = \frac{|\mathrm{Red}(\KG)|}{|\KG|}\;\in[0,1].
\end{equation}
High \srr indicates a bloated graph that raises retrieval cost (\rpc,
\cref{sec:metrics-layer3}) without improving coverage; it is computable without
an ontology.

\medskip\noindent\textbf{Benchmark metrics (require $\refKG$).}

When a reference KG $\refKG$ is available (e.g., PathQuestion~\cite{zhou2018interpretable}, MetaQA~\cite{zhang2018variational}), the following metrics from~\cite{TKG-Amdouni2026} provide gold-anchored evaluation and calibrate the ground-truth-free metrics above.
Let $\KG_{\mathrm{crt}}=(\KG\cap\refKG)|_{\Pi_{\KG}}$ be the correct subset of $\KG$ and $\KG_{\mathrm{cpt}}=(\KG\cap\refKG)|_{\Pi_{\refKG}}$ the covered subset of $\refKG$.

\paragraph{Correctness (\CORm, $\mcorrect$)~\cite{TKG-Amdouni2026}.}
\begin{equation}
  \mcorrect = 1 - \frac{|\KG\setminus\KG_{\mathrm{crt}}|}{|\KG|}\;\in[0,1].
\end{equation}

\paragraph{Completeness (\CMPm, $\mcomplete$)~\cite{TKG-Amdouni2026}.}
\begin{equation}
  \mcomplete = 1 - \frac{|\refKG\setminus\KG_{\mathrm{cpt}}|}{|\refKG|}\;\in[0,1].
\end{equation}
Low \CMPm is the gold-anchored counterpart of low \oc: both signal
structural incompleteness, but \CMPm is measured against gold triples and
\oc against the ontology schema.
In the \tkg lifecycle~\cite{TKG-Amdouni2026}, low \CMPm is a direct
trigger for a \tkgterm{KG Update}.

\paragraph{Remediation note (Validation).} Low Validation metrics call for \emph{graph-and-schema levers} that touch the graph and its schema but not the retrieval algorithm: improve entity resolution and canonicalization (high \dnr), prune or merge duplicates (high \srr), repair contradicting triples or axioms (low \LCm), and, importantly, refine, loosen, or extend the ontology, or acquire targeted documents, when \oc or \CMPm is low.
Ontology refinement is the natural Validation-stage lever, available only when a schema exists; without one, the resolution-based fixes above carry the load.
These are the graph-and-schema levers of the remediation map (\cref{sec:remediation}). 
\subsection{KG Usage Metrics}
\label{sec:metrics-layer3}

The \tkgterm{KG Usage} phase queries the validated KG via graph-based retrieval and inference.
We first fix the object these metrics operate on.
For a query $q$ over the validated KG $\KG$ with vertex set $V$, the retrieval function induces a \emph{retrieved subgraph} $\KG_S\subseteq\KG$ with vertex set $V_S\subseteq V$.
A path ``in the subgraph'' uses only edges in $\KG_S$.
All intrinsic metrics below are ground-truth-free and computable before or during generation; gold-requiring metrics are collected at the end.

\emph{Core question:} does retrieval set up a correct answer?
The signal to watch first is Entity Retrieval Coverage (\erc), with Reasoning Readiness (\rrs) close behind; the answer-side metrics then check faithfulness.

We sort the usage metrics into two roles and label each group below accordingly. A \emph{per-query predictor} takes a graded value on a single query, so it can flag or triage that query before the answer is judged, and it can also be aggregated as the averaged score over a set of queries. A \emph{global outcome metric} instead summarizes end-task success over a \emph{set} of queries, as a rate or ratio, and serves as an indicator.
This predictor/outcome split is orthogonal to the primary/derived distinction of \cref{tab:corederived}: the former concerns what a metric measures (a single query versus a set), the latter whether it is an independent signal or a definitional complement.

\subsubsection{Query and Answer Entity Resolution}

Usage metrics compare \emph{entities} of the query or answer against $\KG_S$.
Resolution proceeds in two stages: an \emph{extraction} step turns free text into candidate entity mentions (an NER model or a noun-chunk tokenizer), and a \emph{matching} step resolves each mention to a graph node by significant-token overlap against node labels.
Let \(M_q\) be the set of surface mentions
extracted from query \(q\), and let
\(\gamma_q : M_q \to V \cup \{\bot\}\) be the grounding function, where
\(\bot\) denotes an unresolved mention. We define the grounded query-node set as
\[
V_q^g = \{\gamma_q(m) : m \in M_q,\ \gamma_q(m) \neq \bot\} \subseteq V .
\]
The query-side denominator remains the number of extracted surface mentions
\(|M_q|\), so that missing or unresolved mentions are penalized. 
Thus \qgr, \erc, and \rrs measure, respectively, the fraction of surface mentions
grounded to the KG, the fraction whose grounded nodes are retrieved into the
subgraph, and the fraction of query-mention pairs whose grounded nodes are
connected there.
The retrieved subgraph has node set \(V_S \subseteq V\). All intersections such
as \(V_q^g \cap V_S\) are therefore intersections between KG node identifiers,
not between text mentions and graph nodes.
Over-segmented multi-word mentions are then coalesced into the largest matched entity, so each real-world entity contributes a single node.
All query-side metrics are measured over the same mention universe \(M_q\):
QGR and ERC use \(|M_q|\) directly, while RRS uses query-mention pairs through
\(\binom{|M_q|}{2}\). Thus unresolved or unretrieved mentions are penalized
consistently across scope, coverage, and connectivity.

\emph{Convention for degenerate inputs.} Throughout this section, a ratio with an empty denominator is reported as N/A rather than assigned a numeric value: \qgr, \erc, and \rrs are N/A when $|M_q|=0$ (no entity mention could be extracted from the query); \erp and \aur are N/A when $|V_S|=0$ (empty retrieval); and \agr is N/A when $|M_a|=0$ (no entity mention could be extracted from the answer).
For \rhdm, queries whose grounded entities are disconnected in $\KG_S$ (no finite $\ell(q)$) are collected in a separate $\ell=\infty$ bucket, so the reported distribution sums to one over $\{0,1,2,\dots,\infty\}$.
These degenerate cases are themselves diagnostic (an empty $V_q^{\mathrm{g}}$ or $V_S$ signals an out-of-scope query or a retrieval miss) and are reported as counts alongside the metric.

\subsubsection{Retrieval-time Metrics}

These characterize the retrieved subgraph \emph{before} the LLM is invoked. They are all per-query predictors.

\paragraph{Query Grounding Rate ($\qgr$) \textbf{[TRIAGE]}.}
Fraction of extracted query mentions that exist as graph nodes at all, a KG-coverage signal distinct from retrieval quality:
\begin{equation}
  \qgr = \frac{|V_q^{\mathrm{g}}|}{|M_q|}\;\in[0,1].
\end{equation}

\paragraph{Entity Retrieval Coverage ($\erc$) \textbf{[TRIAGE]}.}
Fraction of \emph{all} extracted query mentions whose grounded nodes are present in the retrieved subgraph:
\begin{equation}
  \erc = \frac{|V_q^{\mathrm{g}}\cap V_S|}{|M_q|}\;\in[0,1].
\end{equation}
Sharing the denominator $|M_q|$ with \qgr gives $\erc\le\qgr$ always; the gap
$\qgr-\erc$ is exactly the retrieval failure (grounded entities not fetched).
Because $M_q$ is extracted from the query surface, \erc never inspects the answer: it is disjoint from answer-side outcomes such as \ctxcov, so using \erc to predict them is not circular by construction.

\paragraph{Entity Retrieval Precision ($\erp$) \textbf{[TRIAGE]}.}
The precision complement of \erc: are retrieved nodes mostly query-relevant, or bloated by neighborhood expansion?
\begin{equation}
  \erp = \frac{|V_q^{\mathrm{g}}\cap V_S|}{|V_S|}\;\in[0,1].
\end{equation}
Because the denominator $|V_S|$ counts \emph{all} retrieved nodes, typically far larger than the handful of query anchors, \erp is low in absolute terms by
construction; it is therefore read \emph{relatively}, comparing retrieval
strategies or configurations on the same queries rather than against a fixed
threshold. Low \erp alongside high \erc indicates over-retrieval: a large context
with few anchors to the question.

\paragraph{Reasoning Readiness Score ($\rrs$) \textbf{[TRIAGE]}.}
Whether the subgraph is connected enough to support reasoning across the
query's entities: the fraction of query-entity pairs joined by a path in
$\KG_S$, with a convention for the single-entity case where the pair count is
zero:
\begin{equation}
  \rrs =
  \begin{cases}
    \dfrac{|\{\{e_i,e_j\}\subseteq V_q^{\mathrm{g}} :
      \exists\,\mathrm{path}(e_i,e_j)\ \text{in}\ \KG_S\}|}{\binom{|M_q|}{2}},
      & |M_q|\ge 2,\\[1.2em]
    \mathbb{I}\big[V_q^{\mathrm{g}}\cap V_S \neq \emptyset\big],
      & |M_q|=1,
  \end{cases}
\end{equation}
By construction, \(\rrs\in[0,1]\).
The denominator counts pairs over \emph{all} extracted query mentions, so an
off-graph entity is penalized.
For a single-entity query, readiness reduces to whether that entity is grounded
and retrieved ($\rrs=1$) or not ($\rrs=0$).
Because this denominator also charges \rrs for entities that were never grounded
or retrieved, a low \rrs can reflect a grounding or retrieval gap rather than a
genuine connectivity failure.
\rrs is therefore read as a \emph{pure} connectivity signal only conditional on
high \erc: the diagnostic ``high \erc, low \rrs'' connectivity gap of
\cref{sec:remediation} is exactly the regime in which the entities are present
yet disconnected.

A more localized variant could replace the denominator by the number of grounded, retrieved query-entity pairs, e.g., pairs in $V_q^g \cap V_S$, but we keep the global denominator over $M_q$ because RRS is used in the diagnostic chain as an end-to-end readiness signal: it should decrease not only when retrieved entities are disconnected, but also when grounding or retrieval has already failed.

\paragraph{Retrieval Path Cost ($\rpc$) \textbf{[TRIAGE]}.}
The intrinsic metrics above say nothing about \emph{cost}, yet the central
question of \cref{sec:intro} includes the price of traversal.
\rpc captures it as the retrieval effort expended per query, measured as the
number of nodes expanded during subgraph construction (a hardware-independent
proxy) and, optionally, wall-clock path-search time:
\begin{equation}
  \rpc(q) = |\{v\in V : v\ \text{expanded during retrieval for}\ q\}|.
\end{equation}
\rpc makes the coverage/cost trade-off observable: a retrieval strategy may
raise \erc and \rrs only by expanding far more of the graph, and \rpc exposes
that price.
It connects directly to \srr (\cref{sec:metrics-layer2}), since a redundant
graph inflates expansion, and to the traversal-budget failure mode of
\cref{sec:related}.

\subsubsection{Answer-time Metrics}

These measure faithfulness of the generated answer to the retrieved subgraph.
Let \(M_a\) be the set of surface mentions extracted from the generated answer
\(A\), and let \(\gamma_a : M_a \to V \cup \{\bot\}\) be the answer grounding
function. We define
\[
V_a^g = \{\gamma_a(m) : m \in M_a,\ \gamma_a(m) \neq \bot\}.
\]
\agr, \aur, and \arf are per-query predictors; \rhdm is a set-level metric and so reads as a diagnostic rather than a single-query signal.

\paragraph{Answer Grounding Rate ($\agr$) \textbf{[TRIAGE]}.}
Fraction of answer entities present in the retrieved context, detecting
hallucinated entities never retrieved:
\begin{equation}
  \agr = \frac{|V_a^g\cap V_S|}{|M_a|}\;\in[0,1].
\end{equation}

\paragraph{Answer Utilization Rate ($\aur$) \textbf{[TRIAGE]}.}
The recall complement of \agr, how much of the retrieved context the answer
actually used:
\begin{equation}
  \aur = \frac{|V_a^g\cap V_S|}{|V_S|}\;\in[0,1].
\end{equation}
As with \erp, the denominator $|V_S|$ makes \aur structurally low in absolute
terms, so it is interpreted relatively across queries or configurations rather
than against a fixed threshold. High \agr with low \aur suggests a correct answer
drawn from a narrow slice of available evidence.

\paragraph{Answer Reasoning Faithfulness ($\arf$) \textbf{[TRIAGE]}.}
\arf measures \emph{relational} faithfulness: when the answer discusses two
entities the KG connects, does it state their relationship correctly?
This isolates relational hallucination, which entity-level \agr cannot see.
Rather than parsing the answer for explicit relational claims (unreliable,
because extracting relations from generated text propagates the same errors that
afflict any extractor and compounds them at each
stage~\cite{cui2018neuralopenie,llmoie2023uncertainty}), \arf uses the
retrieved triples as the reference set of relations the answer \emph{could}
assert.
Let $C_q$ be the retrieved \emph{co-mentioned} triples whose endpoints are both grounded in
the answer; \arf is the fraction whose relation also appears in the answer text
$A$:
\begin{equation}
  \arf = \frac{|\{(s,p,o)\in C_q : p\subseteq A\}|}{|C_q|}, 
  \quad \\
  C_q=\{(s,p,o)\in \KG_S : s\in V_a^g\cap V_S \wedge o\in V_a^g\cap V_S\},
\end{equation}
where $p\subseteq A$ tests lexically whether the relation label appears in
$A$.
\arf is reported as N/A when $|C_q|=0$ (no connected pair co-mentioned).

\paragraph{Reasoning Hop Depth ($\rhdm$) \textbf{[TRIAGE]}, diagnostic.}
\rhdm characterizes the structural complexity actually exploited, as the
distribution of shortest-path lengths $\ell(q)$ connecting the grounded query
entities in $\KG_S$, 
with \(\ell(q)=0\) for single-entity queries whose grounded entity is retrieved,
and \(\ell(q)=\infty\) when the grounded query mentions are not connected in
\(K_S\) or cannot be evaluated because grounding/retrieval failed.

\begin{equation}
  \rhdm(k) = \frac{|\{q\in\mathcal{Q} : \ell(q)=k\}|}{|\mathcal{Q}|},
  \quad k\in\{0,1,2,\dots,\infty\},
\end{equation}
so that $\sum_k \rhdm(k)=1$ including the disconnected ($\infty$) bucket.
Mass concentrated at $k\le 1$ indicates the pipeline rarely exploits
multi-hop structure, suggesting the KG adds little over flat retrieval for that
query class, while mass at $\infty$ flags queries the subgraph cannot connect at
all. \rhdm is reported as a distribution to preserve diagnostic detail.

\subsubsection{Gold-requiring Usage Metrics}

When gold reasoning paths or gold answers are available (as in PathQuestion and MetaQA), the Usage phase admits gold-anchored metrics that give it the same two-regime treatment as the earlier phases. \gpc and Answer Correctness (Hits@1/F1) are global outcome metrics; RPP/RPR are per-query but, unlike the predictors above, need a gold path.

\paragraph{Gold Path Coverage ($\gpc$) \textbf{[TRIAGE]}.}
Fraction of queries whose gold reasoning path $P^*_q$ is fully contained in
the retrieved subgraph:
\begin{equation}
  \gpc = \frac{|\{q\in\mathcal{Q} : P^*_q\subseteq \KG_S\}|}{|\mathcal{Q}|}
  \;\in[0,1].
\end{equation}
\gpc is the gold-anchored counterpart of \rrs/\erc.

\paragraph{Retrieval Path Precision / Recall (RPP / RPR) \textbf{[TRIAGE]}.}
Of the retrieved-subgraph edges, the fraction lying on the gold path
(precision), and of the gold-path edges, the fraction retrieved (recall):
\begin{equation}
  \mathrm{RPP} = \frac{|\KG_S\cap P^*_q|}{|\KG_S|},
  \qquad
  \mathrm{RPR} = \frac{|\KG_S\cap P^*_q|}{|P^*_q|}.
\end{equation}

\paragraph{Answer Correctness.} The ultimate gold-anchored outcome, Hits@1 and F1 over answer sets~\cite{yih2016webqsp}, used in \cref{sec:experiments} as the dependent variable the intrinsic metrics are designed to predict.

\subsubsection{Imported Outcome Metrics}

To validate the intrinsic metrics and close the \tkgterm{KG Update} loop, we import four outcome metrics from recent graph-RAG benchmarks.
They are not \tri contributions; we adopt them as dependent variables. Unlike the intrinsic usage metrics above, three of these (\ctxcov, \rfr, \hhr) require a gold answer or correctness label, not live gold-free signals monitored at deployment; only \kgimpact, built from grounding alone, is gold-free. All four are global outcome metrics: \ctxcov and \kgimpact average a per-query label into a rate, whereas \rfr and \hhr are population-level ratios with no single-query value.

\paragraph{KG Net Impact ($\kgimpact$)~\cite{eval-finhallubench2026}.} Counterfactual contribution of the KG: answer grounding \emph{with} the retrieved triples minus that \emph{without} them.
A drift toward $\kgimpact\le 0$ across a query class signals stale or irrelevant KG content.

\paragraph{Context Coverage ($\ctxcov$)~\cite{eval-reasoningbottleneck2026}.} Fraction of queries whose gold answer is present in the retrieved context, the answer-string analogue of \erc.

\paragraph{Reasoning Failure Rate ($\rfr$)~\cite{eval-reasoningbottleneck2026}.} Fraction of queries where the answer is in context yet the model still answers wrongly, isolating the generation-stage residual from retrieval failure.

\paragraph{Hard Hits Rate ($\hhr$)~\cite{eval-whatbreakskgrag2025}.} $\hhr=\mathrm{Hits@Hard}/\mathrm{Hits@Any}$, robustness under an incomplete KG; complements \CMPm by exposing, at usage time, the queries that KG incompleteness breaks.

\paragraph{Remediation note (Usage).} Only \emph{Usage}-stage breaches implicate the retrieval algorithm itself. Concretely, the signature selects the lever: a retrieval gap (high \qgr, low \erc) calls for a wider seed set or traversal budget; a connectivity gap (high \erc, low \rrs) calls for a retrieval primitive that recovers the connecting path, such as backward or path-based traversal; and high \rpc calls for pruning redundant expansion or a sparser primitive.
A retrieval or connectivity gap that persists once extraction and graph health are adequate, or high \rpc, motivates architectural changes to retrieval and reasoning, for example super-relation reasoning that adds backward traversal and aggregates relational paths~\cite{reknos2025}, logic-aware multi-hop traversal~\cite{hoprag2025}, GNN-based structural retrieval~\cite{mavromatis2024gnn}, RL-optimized graph indexing~\cite{autographr1_2025}, agentic restructuring~\cite{agenticrl2025}, or cross-document construction that adds connective structure~\cite{zhang2025rakg}. A low \qgr instead marks an out-of-scope query (no KG answer to retrieve), and a generation gap (an ungrounded answer despite adequate \rrs) points to the generator itself, for example prompting it to cite the retrieved evidence or constraining decoding to it.
These are the retrieval levers, not a prescription; the remediation map is described next. 
\subsection{Metric Dependencies and Redundancy}
\label{sec:metrics-depend}

The suite is deliberately broad, and several metrics are related by construction.
We make these relationships explicit (\cref{tab:corederived}) so that a practitioner can distinguish \emph{primary} signals, worth monitoring and triggering remediation on, from \emph{derived} ones that add diagnostic detail.
We do \emph{not} claim statistical independence, which can only be established empirically (\cref{sec:experiments}); the relationships below are definitional or hypothesized.
We group them into three kinds.
\emph{Definitional complements}: precision/recall pairs over the same sets, where one is the dual of the other (\erc and \erp, \agr and \aur), and \qgr, which fixes the query-mention scope underlying \erc and \rrs.
\emph{Aggregates}: \triti and \tfi are weighted functions of other metrics, and \cvr is a per-constraint refinement of \scomp.
\emph{Expected empirical correlations} (to be tested, not assumed): low \tcs is expected to lower \scr, since \scr counts confident triples; high \dnr is expected to lower \rrs, since dead nodes cannot form connecting paths; high \srr is expected to raise \rpc, since redundant structure inflates expansion.
For remediation, the primary metrics are the triggers; the derived metrics help localize \emph{why} a primary metric moved.

\subsection{From Metrics to Remediation}
\label{sec:remediation}

\begin{figure}[t]
  \centering
  \begin{tikzpicture}[
    font=\footnotesize, >=Latex, node distance=4mm and 14mm,
    chk/.style={draw=blue!55, fill=blue!5, rounded corners, align=center,
      minimum width=22mm, minimum height=7mm},
    sig/.style={draw=orange!70!black, fill=orange!8, rounded corners, align=left,
      text width=52mm, minimum height=7mm, inner sep=3pt},
    ok/.style={draw=green!50!black, fill=green!10, rounded corners, align=center,
      minimum width=22mm, minimum height=7mm},
  ]
    \node[chk] (q) {$\qgr<\theta$?};
    \node[sig, right=of q] (s1) {\textbf{out-of-scope/grounding failure}: entities not grounded in the KG};
    \node[chk, below=of q] (e) {$\erc<\theta$?};
    \node[sig, right=of e] (s2) {\textbf{retrieval gap}: in KG, not retrieved};
    \node[chk, below=of e] (r) {$\rrs<\theta$?};
    \node[sig, right=of r] (s3) {\textbf{connectivity gap}: retrieved, not connected};
    \node[chk, below=of r] (a) {$\arf<\theta$?};
    \node[sig, right=of a] (s4) {\textbf{relational gap}: connected, relation wrong};
    \node[chk, below=of a] (g) {$\agr<\theta$?};
    \node[sig, right=of g] (s5) {\textbf{generation gap}: answer not grounded};
    \node[ok, below=of g] (okn) {answer supported};
    \draw[->] (q) -- node[above]{yes} (s1);
    \draw[->] (e) -- node[above]{yes} (s2);
    \draw[->] (r) -- node[above]{yes} (s3);
    \draw[->] (a) -- node[above]{yes} (s4);
    \draw[->] (g) -- node[above]{yes} (s5);
    \draw[->] (q) -- node[left]{no} (e);
    \draw[->] (e) -- node[left]{no} (r);
    \draw[->] (r) -- node[left]{no} (a);
    \draw[->] (a) -- node[left]{no} (g);
    \draw[->] (g) -- node[left]{no} (okn);
  \end{tikzpicture}
  \caption{The usage-stage diagnostic chain as a decision cascade. Reading the
  metrics in order, the first one below its threshold $\theta$ names the failure's
  \emph{signature}; if all pass, the answer is structurally supported. Each
  signature then maps to the levers that can remedy it; the chain localizes the
  failing \emph{condition}, whose root cause may lie upstream.}
  \label{fig:chain}
\end{figure}
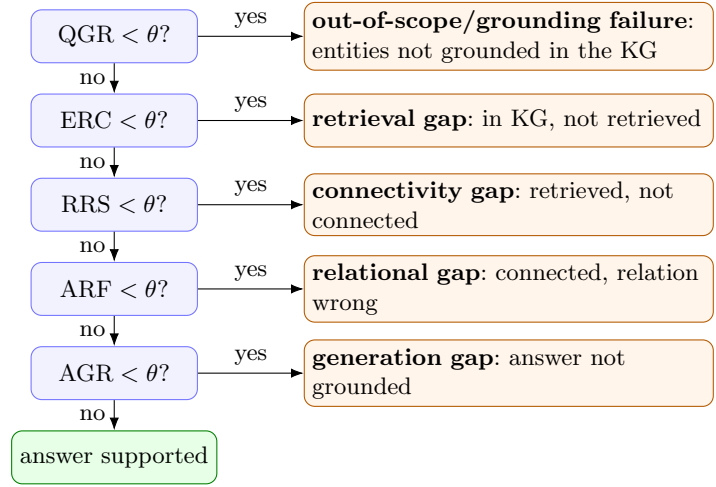

\paragraph{The usage-stage diagnostic chain.} At inference time, a correct graph-grounded answer requires a chain of conditions to hold in sequence (\cref{fig:chain}), each one necessary for the next, and each observable without gold annotations.
The query entities must exist in the KG (\qgr), be retrieved into the subgraph (\erc), be connected there by a path (\rrs), have their relation stated correctly by the answer (\arf), and appear in an answer grounded on the retrieved evidence (\agr).
Each condition presupposes the previous one: a query entity cannot be retrieved unless it is in the KG, cannot be connected unless it is retrieved, and so on.
For the first two links this is exact, $\erc \le \qgr$ by construction; for the later links it is a logical precondition rather than a numeric bound.
This ordering is what makes the chain diagnostic: reading the metrics in sequence, the \emph{first} link that falls below its threshold $\theta$ localizes the failure to a single condition.
We call the resulting label the failure's \emph{signature}: out-of-scope or grounding failure (low \qgr), retrieval gap (high \qgr, low \erc), connectivity gap (high \erc, low \rrs), relational gap (high \rrs, low \arf), and generation gap (high \arf, low \agr).
The first three links are computable before the LLM is invoked, enabling pre-generation triage; the last two require the produced answer and so apply post hoc.
Each link is a necessary, observable proxy rather than a guarantee: a satisfied link is required for success but does not certify it, since a structural signal such as \rrs cannot distinguish a correct connecting path from a spurious one.
The thresholds $\theta$ separating ``low'' from ``adequate'' are not assumed; \cref{sec:experiments} specifies how to estimate them empirically.

\paragraph{What the chain does and does not claim.} The chain is a usage-stage detector: it operates on a query and the validated KG, and it localizes \emph{which} condition failed, not \emph{why}.
The cause of a broken link may lie upstream, in extraction or graph construction, so the build- and validation-time metrics (\cref{sec:metrics-layer1}, \cref{sec:metrics-layer2}) are the candidate explanations, and the localization test in Section~\ref{sec:protocol} examines whether they co-move with chain failures.

\paragraph{From a signature to remediation.}
Diagnosis is the contribution; what to do about a signature is a map from the failing stage to the levers that can act on it, not a fixed order.
Each stage has characteristic levers:
\begin{enumerate}
  \item \textbf{Extraction levers.} Low \tcs, \scr, \ec, \xsc, or \scomp/\cvr points to extraction, addressable by prompt refinement, lower decoding temperature, rebalanced few-shot demonstrations, re-chunking, or targeted re-extraction of low-coverage regions, with no change to graph structure or retrieval.
  \item \textbf{Graph and schema levers.} Low \oc, high \dnr, high \srr, low $\mlc$, or low $\mcomplete$ points to the assembled graph or its schema, addressable by improved entity resolution and canonicalization, duplicate pruning, contradiction repair, ontology refinement or extension, and document acquisition, all of which touch the KG but not the retrieval algorithm.
  \item \textbf{Retrieval levers.} A retrieval or connectivity gap, or high \rpc, that persists once extraction and graph health are adequate points to the retrieval and reasoning algorithm; example directions are given in the Usage remediation note above.
\end{enumerate}
We deliberately do not rank these levers by cost: re-extraction reprocesses the whole corpus, a graph fix reprocesses the assembled graph, and a retrieval change may be a configuration swap or a new method, so their relative cost is deployment-specific rather than universal.
Which lever is cheapest and most effective for a class of failures is an empirical question for the validation protocol in Section~\ref{sec:protocol} and future remediation studies.
What \tri adds over a lifecycle that triggers an undifferentiated update is this targeting: the diagnosis names the stage and the levers that can address it.

\paragraph{Closing the loop.} Each lever set corresponds to re-entering a \tkgterm{KG Update} cycle at a different depth: extraction levers re-run extraction, graph-and-schema levers revise the assembled graph or ontology, and retrieval levers alter the retrieval algorithm.
A KG Update modifies the shared graph, so it must be justified by population behavior rather than a single query: the build- and validation-time metrics are already graph-level, whereas the usage-stage metrics are per-query and must be aggregated before they can trigger an update.
For a per-query metric $m$ and a query window $W$, an update fires when $\mathrm{agg}_{q\in W}\, m(q) < \theta$, with $\mathrm{agg}$ a mean, a low quantile, or the rate of queries failing the threshold; a single low $m(q)$ drives only per-query triage, a fallback or a warning, not a change to the graph.
A persistent aggregate breach then triggers an update targeted at the diagnosed stage, which operationalizes the \tkg lifecycle dimension~\cite{TKG-Amdouni2026} as a control loop for automated graph-RAG deployments.

\section{Recommended Evaluation Protocol}
\label{sec:experiments}

This section makes one empirical point and frames the rest.
Because \tri's deployed instrument is gold-free, its evaluation is a question of \emph{construct validity} resting on three claims: that the suite is \emph{computable} gold-free on real graphs, \emph{predictive} of downstream failure, and \emph{localizing} of where a failure originates.
Two of these we establish here, on the \emph{usage} metrics.
\emph{Computability} we show by construction, computing the full gold-free usage suite across five retrievers and $480$ query-by-retriever observations on a real KG with no gold annotation at scoring time (\cref{sec:poc}).
The proof of concept provides preliminary evidence of predictive validity: the structural metrics indicate whether a faithful answer’s evidence is present in the retrieved subgraph, when the reader may often produce an answer that \textit{seems} correct from parametric memory, although retrieval has failed.
In that case end-answer accuracy would certify a graph our metrics correctly flag as deficient; this silent-success residual is a concrete case for \tri's design, scoring each stage gold-free and \emph{before} the answer rather than trusting end-to-end accuracy.
The \emph{localizing} claim and threshold calibration need larger, more varied benchmarks; rather than run them here we specify them as a falsification protocol (\cref{sec:protocol}) that a later instantiation can execute to test, not assume, them.
The \emph{extraction} metrics, which need a document-grounded corpus, are left to that protocol's extensions.

\subsection{Proof of concept}
\label{sec:poc}

We instantiate the usage stage on \emph{PathQuestion}~\cite{zhou2018interpretable} over its 3-hop knowledge base (PQ-3H: 1,836 entities, 2,839 triples). Because natural PathQuestion items name a single topic entity, leaving \erc and \rrs saturated at $1$, we build a \emph{distance-controlled probe set}: 108 synthetic questions (12 per category) naming two or more KB entities at controlled distances. Each category stresses one link of the diagnostic chain: off-topic questions contain only non-KB entities, semi questions contain one KB entity and one fictional entity, pair-k questions contain two entities at undirected distance k = 1,...,4, disc questions contain two entities in different components, and triple/quad questions contain three or four entities on a connected path.

The 108 questions are designed for diagnostic coverage rather than as a natural query distribution. The 12 off-topic questions test the out-of-scope link of the chain, but they are excluded from the predictive-validity analysis because their gold answers are undefined with respect to the KG. The predictive analysis therefore uses 96 distinct questions. Since each question is run through five retrievers, the pooled analysis contains 480 question-by-retriever observations. These 480 observations should not be read as 480 independent questions; they are five retrieval views of the same 96-question probe.
Entities are sampled at the target structure, verbalized into a fixed template (``how are $X$ and $Y$ related?''), and re-extracted into $M_q$ and grounded into $V_q^g$ by the deployment pipeline; answers come from an LLM reader (\texttt{mistral-small-3.2}).
The structural metrics need no gold answer, and the probe is reproducible from the KB alone.

\textbf{Retrieval configurations used as a diagnostic testbed.}
We do not evaluate these retrievers as systems in their own right. Instead, we use five retrieval configurations as a representative diagnostic testbed: they induce different coverage, connectivity, and cost profiles, allowing us to test whether the proposed structural metrics track evidence availability. Each configuration is a lightweight re-implementation following an existing approach’s core principle rather than an exact reproduction.
The first four share a common $n$-hop beam expansion and differ only in their seed entities: \emph{gold-entity} (the gold topic entity, an oracle upper bound on linking), \emph{extracted-entity} (its deployable counterpart, from LLM-extracted, graph-linked entities), and, following LightRAG's dual-level retrieval~\cite{LightRAG-Guo2024}, \emph{light-entity} (\emph{local} mode: top-$k$ entities most similar to the question) and \emph{light-relation} (\emph{global} mode: endpoints of the top-$k$ most similar triples).
The fifth, \emph{path}, follows PathRAG~\cite{chen2026pathrag}: it forgoes beam expansion and extracts flow-pruned relational paths between anchors.

\paragraph{Predictive validity: structural metrics track context coverage.}
On the PQ-3H probe set we ask whether the structural metrics are associated with the framework’s two gold-anchored outcome metrics (\cref{sec:metrics}).
For each question-retriever pair, we store the retrieved subgraph and the generated answer once. We then compute CC and Hits@1 offline from these stored outputs, without rerunning retrieval or generation: 
we pool all five retrieval configurations into one population of 480 question-by-retriever observations (off-topic excluded), and for each we score two outcomes against the probe’s gold answers: Context Coverage (CC: all gold-answer entities lie in the retrieved subgraph, a reader-free outcome) and Answer Correctness (Hits@1: a gold answer appears in the generated answer). The predictors ERC and RRS are computed without gold annotations; gold answers are used only afterwards to compute CC and Hits@1 as validation outcomes.
We then stratify the population by entity coverage (\erc) and by connectivity (\rrs).
\Cref{tab:poc-pred} reports both outcomes per stratum.

\begin{center}
\fbox{%
\begin{minipage}{0.92\linewidth}
\textbf{Example: why answer correctness can hide a retrieval failure.}
Consider a query ``How are \(X\) and \(Y\) related?'', whose gold answer requires an intermediate entity \(Z\). A retriever may return both query entities \(X\) and \(Y\), giving \(ERC = 1\), but fail to retrieve a connecting path between them, giving \(RRS = 0\). If \(Z\) is absent from the retrieved subgraph, then \(CC = 0\), because the evidence needed for a faithful answer is missing. A strong reader may nevertheless answer with \(Z\) from parametric memory, yielding \(Hits@1 = 1\). In such a case, answer correctness marks the run as successful, whereas the structural metrics flag that the retrieved evidence was insufficient.
\end{minipage}%
}
\end{center}

\textbf{In this controlled probe, the structural metrics strongly separate cases where the gold evidence is present from cases where it is missing.}
Reader-free \ctxcov rises with \erc and jumps sharply once retrieval is complete, from $\mathbf{44.3\%}$ when $\erc<0.5$ (and a near-flat $\mathbf{45.5\%}$ for partial coverage) to $\mathbf{98.4\%}$ at $\erc=1$ (a $54$-point gap), and the Youden-optimal operating point for predicting low \ctxcov sits at $\theta_{\erc}{=}0.67$ (Youden $J{=}0.67$).
Connectivity tells the same story on its own axis: a disconnected retrieval ($\rrs{=}0$) covers the gold answer only $\mathbf{54.8\%}$ of the time against $\mathbf{93.2\%}$ when $\rrs>0$.
This suggests, on the PQ-3H probe and across retrieval configurations, that the structural metrics can indicate whether the evidence a faithful answer needs is present.

\textbf{The reader masks the gap, which is why the structural metrics are scored independently.}
Reader Answer Correctness (Hits@1) tracks the same metrics but far more weakly: it climbs only from $\mathbf{77.1\%}$ at $\erc<0.5$ to $\mathbf{91.4\%}$ at full coverage (a $14$-point gap), and stays near $80\%$ even where retrieval is structurally inadequate.
The reader is answering many low-coverage questions from parametric memory rather than from the retrieved subgraph, the silent-success residual of \cref{sec:related}, here measured directly: a correct answer co-occurs with broken retrieval, so end-answer accuracy alone would certify a graph the structural metrics correctly flag as deficient.
This is the concrete motivation for \tri scoring retrieval adequacy \emph{before} and \emph{independently of} the answer.
It also bounds what this probe can show: because the reader floor is high, a clean Hits@1 threshold needs harder questions and a weaker-prior reader, and a calibrated $\theta$ with significance testing is deferred to the larger benchmarks of the protocol below.

Two construct caveats accompany this probe.
The predictor and outcome are disjoint by construction: \erc scores grounded query mentions while \ctxcov scores \emph{gold-answer} entities, so the separation is not tautological, though the same extraction-and-linking pipeline builds $M_q$, $V_q^g$,  and the retrieval, which can inflate \erc independently of graph quality (a coupling the gold-entity oracle retriever isolates).
The threshold $\theta_{\erc}$ is fit in-sample on these $480$ pooled, non-independent observations (five retrievers per question), so it is descriptive rather than a validated operating point, with calibrated thresholds and clustered confidence intervals deferred to the protocol below.

\begin{table}[h]
\centering
\caption{Predictive validity on the PQ-3H probe set, pooling all five
retrievers ($480$ question-by-retriever observations, off-topic excluded).
Context Coverage (\ctxcov) is reader-free (all gold-answer entities in the
retrieved subgraph); Answer Correctness (Hits@1) is the reader outcome. Both rise
with entity coverage (\erc) and connectivity (\rrs), but \ctxcov separates far
more sharply: the structural metrics indicate whether the evidence is present in this controlled probe,
while the strong reader partly masks the gap at answer time (the parametric
silent-success residual). The Youden-optimal threshold for predicting low
\ctxcov is $\theta_{\erc}{=}0.67$. Boldface marks the values discussed in the
text. The table is descriptive: confidence intervals are not reported here because the pooled observations are not independent, with each question appearing under five retrievers; calibrated clustered uncertainty estimates are deferred to the full protocol in Section~\ref{sec:protocol}.}
\label{tab:poc-pred}
\small
\setlength{\tabcolsep}{6pt}
\begin{tabular}{lccc}
\toprule
Stratum & $n$ & \ctxcov (\%) & Hits@1 (\%) \\
\midrule
\multicolumn{4}{l}{\emph{by entity coverage} (\erc)} \\
\quad $\erc<0.5$        & 70  & \textbf{44.3} & \textbf{77.1} \\
\quad $0.5\le\erc<1$    & 165 & \textbf{45.5} & 84.2 \\
\quad $\erc=1$          & 245 & \textbf{98.4} & \textbf{91.4} \\
\midrule
\multicolumn{4}{l}{\emph{by connectivity} (\rrs)} \\
\quad $\rrs=0$          & 261 & \textbf{54.8} & 82.8 \\
\quad $\rrs>0$          & 219 & \textbf{93.2} & 91.8 \\
\bottomrule
\end{tabular}
\end{table}

\FloatBarrier
\subsection{A falsification protocol for trust metrics}
\label{sec:protocol}

The proof of concept above is deliberately small and structurally controlled. It is sufficient to show that the usage metrics are computable and that, in one setting, they separate evidence-present from evidence-missing cases. It is not sufficient to validate thresholds, prove localization, or demonstrate operational triage utility. We therefore treat the remaining claims as hypotheses and specify, for each one, the expected effect, the concrete test, and the outcome that would \emph{refute} it, so that a later instantiation tests the claim rather than searching for support.
All comparisons share one reader; answer quality uses Hits@1 and answer-set F1; confidence intervals use BCa bootstrap resampling~\cite{efron1987bca,efron1994bootstrap} and paired comparisons a paired bootstrap, and a claim holds only when its effect lies in the predicted direction with a confidence interval excluding the null.

\paragraph{Predictive validity (full study).}
\emph{Expected:} Hits@1 increases monotonically across \erc (and separately \rrs) quartiles, with a sharp drop below a threshold $\theta$.
\emph{Test:} stratify queries by quartile on \emph{WebQSP}~\cite{yih2016webqsp} (4{,}737 natural questions over Freebase, 1--2 hop) and report per-quartile Hits@1 with paired-bootstrap significance for the Q4--Q1 gap; estimate $\theta$ by the Youden index for predicting answer failure; repeat per hop depth on \emph{PathQuestion}~\cite{zhou2018interpretable} ($\approx$2.6K synthetic multi-hop questions with gold reasoning paths) and MetaQA~\cite{zhang2018variational}.
\emph{Falsified if} the Q4--Q1 gap is null or the ordering is non-monotone; passing extends the proof of concept to natural questions and measures the parametric silent-success residual of \cref{sec:related}, where the reader succeeds despite low \erc.

\paragraph{Localization.}
\emph{Expected:} low upstream graph-health metrics (e.g.\ \tcs at Implementation, \CMPm at Validation) co-move with low downstream \erc/\rrs, and the five diagnostic signatures of \cref{sec:remediation} (out-of-scope, retrieval gap, connectivity gap, relational gap, generation gap) are mutually distinguishable by their metric profiles.
\emph{Test:} report per-query Spearman $\rho$ between each upstream metric and \erc/\rrs, with confidence bands and stratification by query type, together with the fraction of failures in each signature; this is an alignment study, not a causal-propagation claim.
\emph{Falsified if} the correlations are null or the signatures are not separable, in particular if the relational and generation gaps fail to isolate failures that survive the structural links (high \erc/\rrs).

\paragraph{Utility.}
\emph{Expected:} routing low-\erc/\rrs queries before generation, to a fallback retrieval strategy or a low-confidence warning, improves end-task reliability over no triage.
\emph{Test:} compare the same reader over the same KG with and without \erc/\rrs pre-inference triage, reporting Hits@1/F1 and the fraction of silent failures averted.
\emph{Falsified if} triage yields no reliability gain at any operating point.

Two threads are left to future work.
Triple-level metrics that depend on human judgment (e.g.\ fidelity calibration) require double annotation against the source, labelling each triple \emph{Supported}, \emph{Unsupported}, or \emph{Contradicted} with a third annotator adjudicating, and reporting Cohen's $\kappa$~\cite{cohen1960kappa} and Krippendorff's $\alpha$~\cite{krippendorff2004contentanalysis}.
Three validations also remain open: the \emph{extraction} metrics on a document-grounded corpus (white-box construction with source text), whether ontology grounding raises structural quality (\oc up, \dnr and \srr down), and the remediation map, namely that the lever indicated by a diagnosed stage resolves a substantial share of failures at a measurable, deployment-specific cost.

\section{Conclusion}
\label{sec:conclusion}

We presented \tri, a stage-aware instrumentation framework that instruments automated, document-grounded graph-RAG pipelines end to end, with construction-stage metrics that apply to any extracted KG, automated or not.
Informed by the \tkg engineering methodology~\cite{TKG-Amdouni2026} and adapted to the automated setting, \tri attaches stage-specific, independently interpretable confidence metrics to three phases (\tkgterm{KG Implementation}, \tkgterm{KG Validation}, and \tkgterm{KG Usage}) without requiring a single composite score and without relying on gold annotations at deployment time.
Its central question is deliberately twofold: not only whether the graph underpinning retrieval can be trusted, but at what retrieval and computational cost it can be used.

\paragraph{Summary of contributions.} Three ideas distinguish \tri beyond the metric definitions themselves.
First, it organizes metrics by two computability axes, the external reference each needs (none, an ontology, or a gold standard) and the visibility of the extraction (white-box or black-box), making explicit which signals survive in gold-free, schema-free, or black-box settings and showing that the core assessment degrades gracefully rather than failing when conditions are not ideal; the gold-requiring metrics sit apart as offline calibration yardsticks rather than deployment signals.
Second, it separates primary signals from derived ones through a dependency analysis, yielding a minimal set worth monitoring.
Third, and most distinctively, it shows that the usage-stage metrics form a diagnostic chain whose first broken link localizes a failure, and that this localization maps to the stage levers that can remedy it: extraction, graph-and-schema, or retrieval.
Targeting the update to the diagnosed stage is what separates \tri from a lifecycle that triggers an undifferentiated update; the relative cost of the levers is deployment-specific and left to empirical measurement (\cref{sec:experiments}).
A usage-stage proof of concept gives preliminary, single-probe evidence for this diagnostic claim: a capable reader can answer from its parametric knowledge when retrieval has missed the needed evidence, so retrieval adequacy must be scored before and independently of the answer (\cref{sec:experiments}).

\paragraph{TRIAGE as one iteration of the TKG lifecycle.} \tri covers the \tkgterm{methodology dimension} of \tkg for one KG version.
Metric breaches, such as a rising \dnr or low aggregate \erc/\rrs over a query class, or rising \rpc, provide principled triggers for a new \tkgterm{KG Update} cycle, targeted at the diagnosed stage.
This closes the loop with the \tkg \tkgterm{lifecycle dimension} and points toward continuous, monitored KG evolution in operational deployments.

\paragraph{Limitations.} Several limitations should be acknowledged.
First, validation is still partial: a usage-stage proof of concept exercises the predictive claim, but the full suite, the remediation map, and the cross-stage links are specified rather than tested at scale; \cref{sec:experiments} gives the protocol for the remaining validation.
Second, \tcs relies on LLM token log-probabilities as a confidence proxy, which are known to be imperfectly calibrated for factual correctness~\cite{kadavath2022know,jiang2021when}; a miscalibrated \tcs may flag correct triples or pass incorrect ones.
Third, the \tkgterm{KG Validation} expert step is specified methodologically but its cost and reliability at scale remain open.
Fourth, the schema-dependent metrics (\oc, \scomp, \LCm, \CMPm) assume an available ontology; without one, assessment relies on the schema-free subset, as \cref{sec:metrics} makes explicit.
Fifth, \arf uses lexical relation matching, which may miss paraphrased or implicit relational claims.
Sixth, the separation of primary from derived metrics is definitional rather than statistical: genuine non-redundancy can only be established empirically (\cref{sec:experiments}), and some primaries may prove correlated in practice.
Seventh, the usage-stage links are necessary but not sufficient conditions: a structural signal such as \rrs cannot distinguish a correct connecting path from a spurious one, so a satisfied chain certifies retrievability, not answer correctness.
Eighth, the usage metrics depend on reliable query- and answer-entity extraction, so errors in $M_q$, $M_a$, or their grounding into $V_q^g$ and $V_a^g$ propagate into \qgr, \erc, \rrs, \agr, \aur, and \arf, and the decision thresholds $\theta$ estimated on a benchmark may not transfer unchanged to deployment.

\paragraph{Future work.} Beyond empirical validation, \tri opens several directions: (i)~extending to multi-hop, multi-entity query decomposition; (ii)~automating \tkgterm{KG Update} triggers from metric thresholds over a rolling query window, with the diagnosis selecting the levers; (iii)~formalizing the \tkgterm{KG Validation} expert step with active learning to reduce annotation cost; (iv)~replacing lexical \arf with an NLI-based relational-faithfulness check; and (v)~surfacing the stage-localized metrics in a per-phase diagnostic interface that maps metric drifts onto the remediation map, enabling continuous, cost-aware KG health monitoring in operational deployments.

\section*{Acknowledgments}

The authors thank Sabrina Chaouche and Emna Amdouni for their help and remarks.

\clearpage  
\bibliographystyle{splncs04}
\bibliography{tsl_references}

@inproceedings{TKG-Amdouni2026,
  author    = {Amdouni, Emna and Mattioli, Lucas and Adjed, Faouzi and
               Awadid, Afef and Gonzalez, Martin and Cantat, Loic and
               Mattioli, Juliette},
  title     = {An End-to-End Trustworthy Knowledge Graph Engineering Methodology},
  booktitle = {16th International Conference on Performance, Safety and Robustness
               in Complex Systems and Applications (PESARO)},
  year      = {2026}
}

@article{GraphRAG-Edge2024,
  author    = {Edge, Darren and Trinh, Ha and Cheng, Newman and Bradley, Joshua and
               Chao, Alex and Mody, Apurva and Truitt, Steven and
               Metropolitansky, Dasha and Ness, Robert Osazuwa and Larson, Jonathan},
  title     = {From Local to Global: {A} Graph {RAG} Approach to
               Query-Focused Summarization},
  journal   = {arXiv preprint arXiv:2404.16130},
  year      = {2024}
}

@article{LightRAG-Guo2024,
  title={Lightrag: Simple and fast retrieval-augmented generation},
  author={Guo, Zirui and Xia, Lianghao and Yu, Yanhua and Ao, Tian and Huang, Chao},
  journal={arXiv preprint arXiv:2410.05779},
  volume={2},
  number={3},
  year={2024}
}

@inproceedings{chen2026pathrag,
  title={Pathrag: Pruning graph-based retrieval augmented generation with relational paths},
  author={Chen, Boyu and Guo, Zirui and Yang, Zidan and Chen, Yuluo and Chen, Junze and Liu, Zhenghao and Shi, Chuan and Yang, Cheng},
  booktitle={Proceedings of the AAAI conference on artificial intelligence},
  volume={40},
  number={36},
  pages={30183--30191},
  year={2026}
}

@article{GCR-Luo2024,
  author    = {Luo, Linhao and Zhao, Zicheng and Haffari, Gholamreza and
               Li, Yuan-Fang and Gong, Chen and Pan, Shirui},
  title     = {Graph-Constrained Reasoning: Faithful Reasoning on Knowledge Graphs
               with Large Language Models},
  journal   = {arXiv preprint arXiv:2410.13080},
  year      = {2024}
}

@inproceedings{zhou2018interpretable,
  title={An interpretable reasoning network for multi-relation question answering},
  author={Zhou, Mantong and Huang, Minlie and Zhu, Xiaoyan},
  booktitle={Proceedings of the 27th international conference on computational linguistics},
  pages={2010--2022},
  year={2018}
}

@inproceedings{zhang2018variational,
  title={Variational reasoning for question answering with knowledge graph},
  author={Zhang, Yuyu and Dai, Hanjun and Kozareva, Zornitsa and Smola, Alexander and Song, Le},
  booktitle={Proceedings of the AAAI conference on artificial intelligence},
  volume={32},
  number={1},
  year={2018}
}

@article{mavromatis2024gnn,
  title={Gnn-rag: Graph neural retrieval for large language model reasoning},
  author={Mavromatis, Costas and Karypis, George},
  journal={arXiv preprint arXiv:2405.20139},
  year={2024}
}

@inproceedings{reimers2019sentencebert,
  title        = {Sentence-BERT: Sentence Embeddings using Siamese BERT-Networks},
  author       = {Reimers, Nils and Gurevych, Iryna},
  booktitle    = {Proceedings of the 2019 Conference on Empirical Methods in Natural Language Processing and the 9th International Joint Conference on Natural Language Processing (EMNLP-IJCNLP)},
  year         = {2019},
  pages        = {3980--3990},
  publisher    = {Association for Computational Linguistics},
  address      = {Hong Kong, China},
  doi          = {10.18653/v1/D19-1410},
  url          = {https://aclanthology.org/D19-1410/}
}

@book{salton1983ir,
  title        = {Introduction to Modern Information Retrieval},
  author       = {Salton, Gerard and McGill, Michael J.},
  year         = {1983},
  publisher    = {McGraw-Hill},
  address      = {New York, NY, USA},
  isbn         = {9780070544840}
}

@inproceedings{stanovsky2016oiebenchmark,
  title        = {Creating a Large Benchmark for Open Information Extraction},
  author       = {Stanovsky, Gabriel and Dagan, Ido},
  booktitle    = {Proceedings of the 2016 Conference on Empirical Methods in Natural Language Processing},
  year         = {2016},
  pages        = {2300--2305},
  publisher    = {Association for Computational Linguistics},
  address      = {Austin, Texas},
  doi          = {10.18653/v1/D16-1252},
  url          = {https://aclanthology.org/D16-1252/}
}

@inproceedings{bhardwaj2019carb,
  title        = {{CaRB}: A Crowdsourced Benchmark for Open {IE}},
  author       = {Bhardwaj, Sangnie and Aggarwal, Samarth and Mausam},
  booktitle    = {Proceedings of the 2019 Conference on Empirical Methods in Natural Language Processing and the 9th International Joint Conference on Natural Language Processing (EMNLP-IJCNLP)},
  year         = {2019},
  pages        = {6262--6267},
  publisher    = {Association for Computational Linguistics},
  address      = {Hong Kong, China},
  doi          = {10.18653/v1/D19-1651},
  url          = {https://aclanthology.org/D19-1651/}
}

@article{kuhn1955hungarian,
  title        = {The Hungarian Method for the Assignment Problem},
  author       = {Kuhn, Harold W.},
  journal      = {Naval Research Logistics Quarterly},
  year         = {1955},
  volume       = {2},
  number       = {1--2},
  pages        = {83--97},
  doi          = {10.1002/nav.3800020109}
}

@inproceedings{bowman2015snli,
  title        = {A Large Annotated Corpus for Learning Natural Language Inference},
  author       = {Bowman, Samuel R. and Angeli, Gabor and Potts, Christopher and Manning, Christopher D.},
  booktitle    = {Proceedings of the 2015 Conference on Empirical Methods in Natural Language Processing},
  year         = {2015},
  pages        = {632--642},
  publisher    = {Association for Computational Linguistics},
  address      = {Lisbon, Portugal},
  doi          = {10.18653/v1/D15-1075},
  url          = {https://aclanthology.org/D15-1075/}
}

@inproceedings{williams2018multinli,
  title        = {A Broad-Coverage Challenge Corpus for Sentence Understanding through Inference},
  author       = {Williams, Adina and Nangia, Nikita and Bowman, Samuel},
  booktitle    = {Proceedings of the 2018 Conference of the North American Chapter of the Association for Computational Linguistics: Human Language Technologies, Volume 1 (Long Papers)},
  year         = {2018},
  pages        = {1112--1122},
  publisher    = {Association for Computational Linguistics},
  address      = {New Orleans, Louisiana},
  doi          = {10.18653/v1/N18-1101},
  url          = {https://aclanthology.org/N18-1101/}
}

@inproceedings{honovich2022true,
  title        = {{TRUE}: Re-evaluating Factual Consistency Evaluation},
  author       = {Honovich, Or and Aharoni, Roee and Herzig, Jonathan and Taitelbaum, Hagai and Kukliansy, Doron and Cohen, Vered and Scialom, Thomas and Szpektor, Idan and Hassidim, Avinatan and Matias, Yossi},
  booktitle    = {Proceedings of the 2022 Conference of the North American Chapter of the Association for Computational Linguistics: Human Language Technologies},
  year         = {2022},
  pages        = {3905--3920},
  publisher    = {Association for Computational Linguistics},
  address      = {Seattle, United States},
  doi          = {10.18653/v1/2022.naacl-main.287},
  url          = {https://aclanthology.org/2022.naacl-main.287/}
}

@article{bengio2003neuralLM,
  title        = {A Neural Probabilistic Language Model},
  author       = {Bengio, Yoshua and Ducharme, R{\'e}jean and Vincent, Pascal and Jauvin, Christian},
  journal      = {Journal of Machine Learning Research},
  year         = {2003},
  volume       = {3},
  pages        = {1137--1155},
  url          = {http://www.jmlr.org/papers/v3/bengio03a.html}
}

@inproceedings{brown2020gpt3,
  title        = {Language Models are Few-Shot Learners},
  author       = {Brown, Tom B. and Mann, Benjamin and Ryder, Nick and Subbiah, Melanie and Kaplan, Jared and Dhariwal, Prafulla and Neelakantan, Arvind and Shyam, Pranav and Sastry, Girish and Askell, Amanda and others},
  booktitle    = {Advances in Neural Information Processing Systems},
  year         = {2020},
  volume       = {33},
  pages        = {1877--1901},
  url          = {https://proceedings.neurips.cc/paper/2020/hash/1457c0d6bfcb4967418bfb8ac142f64a-Abstract.html}
}

@misc{kadavath2022know,
  title        = {Language Models (Mostly) Know What They Know},
  author       = {Kadavath, Saurav and Conerly, Tom and Askell, Amanda and Henighan, Tom and Drain, Dawn and Perez, Ethan and Schiefer, Nicholas and Hatfield-Dodds, Zac and DasSarma, Nova and Tran-Johnson, Eli and others},
  year         = {2022},
  eprint       = {2207.05221},
  archivePrefix= {arXiv},
  primaryClass = {cs.CL},
  doi          = {10.48550/arXiv.2207.05221},
  url          = {https://arxiv.org/abs/2207.05221}
}

@inproceedings{wang2023selfconsistency,
  title        = {Self-Consistency Improves Chain of Thought Reasoning in Language Models},
  author       = {Wang, Xuezhi and Wei, Jason and Schuurmans, Dale and Le, Quoc V. and Chi, Ed H. and Narang, Sharan and Chowdhery, Aakanksha and Zhou, Denny},
  booktitle    = {International Conference on Learning Representations (ICLR)},
  year         = {2023},
  eprint       = {2203.11171},
  archivePrefix= {arXiv},
  primaryClass = {cs.CL},
  doi          = {10.48550/arXiv.2203.11171},
  url          = {https://arxiv.org/abs/2203.11171}
}

@article{jaccard1901,
  title        = {{\'E}tude comparative de la distribution florale dans une portion des Alpes et du Jura},
  author       = {Jaccard, Paul},
  journal      = {Bulletin de la Soci{\'e}t{\'e} Vaudoise des Sciences Naturelles},
  year         = {1901},
  volume       = {37},
  pages        = {547--579},
  url          = {https://commons.wikimedia.org/wiki/File:%C3%89tude_comparative_de_la_distribution_florale_dans_une_portion_des_Alpes_et_du_Jura.pdf}
}

@article{cohen1960kappa,
  title        = {A Coefficient of Agreement for Nominal Scales},
  author       = {Cohen, Jacob},
  journal      = {Educational and Psychological Measurement},
  year         = {1960},
  volume       = {20},
  number       = {1},
  pages        = {37--46},
  doi          = {10.1177/001316446002000104}
}

@book{krippendorff2004contentanalysis,
  title        = {Content Analysis: An Introduction to Its Methodology},
  author       = {Krippendorff, Klaus},
  edition      = {2},
  year         = {2004},
  publisher    = {SAGE Publications},
  address      = {Thousand Oaks, CA}
}

@article{efron1987bca,
  title        = {Better Bootstrap Confidence Intervals},
  author       = {Efron, Bradley},
  journal      = {Journal of the American Statistical Association},
  year         = {1987},
  volume       = {82},
  number       = {397},
  pages        = {171--185},
  doi          = {10.1080/01621459.1987.10478410},
  url          = {https://www.jstor.org/stable/2289144}
}

@book{efron1994bootstrap,
  title        = {An Introduction to the Bootstrap},
  author       = {Efron, Bradley and Tibshirani, Robert J.},
  year         = {1994},
  publisher    = {Chapman and Hall/CRC},
  address      = {New York, NY, USA},
  doi          = {10.1201/9780429246593},
  url          = {https://www.taylorfrancis.com/books/mono/10.1201/9780429246593/introduction-bootstrap-bradley-efron-tibshirani}
}

@inproceedings{banko2007textrunner,
  title        = {Open Information Extraction from the Web},
  author       = {Banko, Michele and Cafarella, Michael J. and Soderland, Stephen and Broadhead, Matt and Etzioni, Oren},
  booktitle    = {Proceedings of the 20th International Joint Conference on Artificial Intelligence (IJCAI)},
  year         = {2007},
  pages        = {2670--2676}
}

@inproceedings{bordes2013transe,
  title     = {Translating Embeddings for Modeling Multi-relational Data},
  author    = {Bordes, Antoine and Usunier, Nicolas and Garcia-Duran, Alberto and Weston, Jason and Yakhnenko, Oksana},
  booktitle = {Advances in Neural Information Processing Systems},
  year      = {2013},
  url       = {https://papers.nips.cc/paper/2013/hash/1cecc7a77928ca8133fa24680a88d2f9-Abstract.html},
  note      = {Introduces the standard link prediction evaluation by ranking corrupted head/tail entities and reports MR and Hits@10.}
}

@inproceedings{trouillon2016complex,
  title     = {Complex Embeddings for Simple Link Prediction},
  author    = {Trouillon, Th{\'e}o and Welbl, Johannes and Riedel, Sebastian and Gaussier, Eric and Bouchard, Guillaume},
  booktitle = {Proceedings of the 33rd International Conference on Machine Learning (ICML)},
  year      = {2016},
  volume    = {48},
  series    = {Proceedings of Machine Learning Research},
  pages     = {2071--2080},
  publisher = {PMLR},
  url       = {https://proceedings.mlr.press/v48/trouillon16.html},
  note      = {Reports standard link prediction metrics including MRR and Hits@K on common benchmarks.}
}

@inproceedings{dettmers2018conve,
  title     = {Convolutional 2D Knowledge Graph Embeddings},
  author    = {Dettmers, Tim and Minervini, Pasquale and Stenetorp, Pontus and Riedel, Sebastian},
  booktitle = {Proceedings of the AAAI Conference on Artificial Intelligence},
  year      = {2018},
  volume    = {32},
  number    = {1},
  pages     = {1811--1818},
  doi       = {10.1609/aaai.v32i1.11573},
  url       = {https://ojs.aaai.org/index.php/AAAI/article/view/11573},
  note      = {Uses the standard ranking protocol for link prediction and reports MRR and Hits@K (commonly Hits@1/3/10).}
}

@inproceedings{sun2019rotate,
  title     = {RotatE: Knowledge Graph Embedding by Relational Rotation in Complex Space},
  author    = {Sun, Zhiqing and Deng, Zhi-Hong and Nie, Jian-Yun and Tang, Jian},
  booktitle = {International Conference on Learning Representations (ICLR)},
  year      = {2019},
  url       = {https://openreview.net/forum?id=HkgEQnRqYQ},
  note      = {A standard KGE paper evaluating link prediction with rank-based metrics such as MRR and Hits@K.}
}

@misc{hoyt2022rankmetrics,
  title        = {A Unified Framework for Rank-based Evaluation Metrics for Link Prediction in Knowledge Graphs},
  author       = {Hoyt, Charles Tapley and Berrendorf, Max and Galkin, Mikhail and Tresp, Volker and Gyori, Benjamin M.},
  year         = {2022},
  howpublished = {arXiv preprint},
  eprint       = {2203.07544},
  archivePrefix= {arXiv},
  primaryClass = {cs.LG},
  url          = {https://arxiv.org/pdf/2203.07544},
  note         = {Surveys and formalizes rank-based metrics for KG link prediction, motivating their use in the absence of explicit negatives.}
}

@inproceedings{toutanova2015observed,
  title     = {Observed Versus Latent Features for Knowledge Base and Text Inference},
  author    = {Toutanova, Kristina and Chen, Danqi},
  booktitle = {Proceedings of the 3rd Workshop on Continuous Vector Space Models and their Compositionality (CVSC)},
  year      = {2015},
  pages     = {57--66},
  publisher = {Association for Computational Linguistics},
  address   = {Beijing, China},
  doi       = {10.18653/v1/W15-4007},
  url       = {https://aclanthology.org/W15-4007/}
}

@inproceedings{wei2022chain,
  title     = {Chain-of-Thought Prompting Elicits Reasoning in Large Language Models},
  author    = {Wei, Jason and Wang, Xuezhi and Schuurmans, Dale and Bosma, Maarten and others},
  booktitle = {Advances in Neural Information Processing Systems},
  year      = {2022}
}

@inproceedings{yao2023react,
  title     = {ReAct: Synergizing Reasoning and Acting in Language Models},
  author    = {Yao, Shunyu and Zhao, Jeffrey and Yu, Dian and others},
  booktitle = {International Conference on Learning Representations},
  year      = {2023}
}

@inproceedings{min2022rethinking,
  title     = {Rethinking the Role of Demonstrations: What Makes In-Context Learning Work?},
  author    = {Min, Sewon and Lyu, Xinxi and Holtzman, Ari and Artetxe, Mikel and
               Lewis, Mike and Hajishirzi, Hannaneh and Zettlemoyer, Luke},
  booktitle = {Proceedings of EMNLP},
  year      = {2022}
}

@inproceedings{zhao2021calibrate,
  title     = {Calibrate Before Use: Improving Few-Shot Performance of Language Models},
  author    = {Zhao, Tony and Wallace, Eric and Feng, Shi and Klein, Dan and Singh, Sameer},
  booktitle = {Proceedings of ICML},
  year      = {2021}
}

@inproceedings{maynez2020faithfulness,
  title     = {On Faithfulness and Factuality in Abstractive Summarization},
  author    = {Maynez, Joshua and Narayan, Shashi and Bohnet, Bernd and McDonald, Ryan},
  booktitle = {Proceedings of ACL},
  year      = {2020}
}

@article{ji2023survey,
  title   = {Survey of Hallucination in Natural Language Generation},
  author  = {Ji, Ziwei and Lee, Nayeon and Frieske, Rita and others},
  journal = {ACM Computing Surveys},
  volume  = {55},
  number  = {12},
  pages   = {1--38},
  year    = {2023},
  doi     = {10.1145/3571730}
}

@article{nickel2015review,
  title   = {A Review of Relational Machine Learning for Knowledge Graphs},
  author  = {Nickel, Maximilian and Murphy, Kevin and Tresp, Volker and Gabrilovich, Evgeniy},
  journal = {Proceedings of the IEEE},
  volume  = {104},
  number  = {1},
  pages   = {11--33},
  year    = {2015},
  doi     = {10.1109/JPROC.2015.2483592}
}

@article{paulheim2017kgquality,
  title   = {Knowledge Graph Refinement: A Survey of Approaches and Evaluation Methods},
  author  = {Paulheim, Heiko},
  journal = {Semantic Web},
  volume  = {8},
  number  = {3},
  pages   = {489--508},
  year    = {2017},
  doi     = {10.3233/SW-160218}
}

@article{jiang2021when,
  title   = {How Can We Know When Language Models Know? On the Calibration of Language Models for Question Answering},
  author  = {Jiang, Zhengbao and Araki, Jun and Ding, Haibo and Neubig, Graham},
  journal = {Transactions of the Association for Computational Linguistics},
  year    = {2021},
  volume  = {9},
  pages   = {962--977},
  doi     = {10.1162/tacl_a_00407},
  url     = {https://direct.mit.edu/tacl/article/doi/10.1162/tacl_a_00407/107277/How-Can-We-Know-When-Language-Models-Know-On-the}
}

@article{shen2015entitylinking,
  title   = {Entity Linking with a Knowledge Base: Issues, Techniques, and Solutions},
  author  = {Shen, Wei and Wang, Jianyong and Han, Jiawei},
  journal = {IEEE Transactions on Knowledge and Data Engineering},
  volume  = {27},
  number  = {2},
  pages   = {443--460},
  year    = {2015},
  doi     = {10.1109/TKDE.2014.2327028},
  url     = {https://www.bibsonomy.org/bibtex/27bedb76858c3a0d0094aa8f9392ce270}
}

@inproceedings{ji2014tackbpedl,
  title     = {Overview of {TAC-KBP}2014 Entity Discovery and Linking Tasks},
  author    = {Ji, Heng and Nothman, Joel and Hachey, Ben},
  booktitle = {Proceedings of the Text Analysis Conference (TAC)},
  year      = {2014},
  url       = {http://blender.cs.illinois.edu/paper/edl2014overview.pdf}
}

@inproceedings{guo2017calibration,
  title     = {On Calibration of Modern Neural Networks},
  author    = {Guo, Chuan and Pleiss, Geoff and Sun, Yu and Weinberger, Kilian Q.},
  booktitle = {Proceedings of the 34th International Conference on Machine Learning (ICML)},
  year      = {2017},
  volume    = {70},
  pages     = {1321--1330},
  url       = {https://arxiv.org/abs/1706.04599}
}

@inproceedings{ruffinelli2020you,
  title     = {You CAN Teach an Old Dog New Tricks! On Training Knowledge Graph Embeddings},
  author    = {Ruffinelli, Daniel and Broscheit, Samuel and Gemulla, Rainer},
  booktitle = {International Conference on Learning Representations (ICLR)},
  year      = {2020},
  url       = {https://openreview.net/forum?id=BkxSmlBFvr}
}

@article{eval-ragvsgraphrag2025,
  title={Rag vs. graphrag: A systematic evaluation and key insights},
  author={Han, Haoyu and Ma, Li and Wang, Yu and Shomer, Harry and Lei, Yongjia and Qi, Zhisheng and Guo, Kai and Hua, Zhigang and Long, Bo and Liu, Hui and others},
  journal={arXiv preprint arXiv:2502.11371},
  year={2025}
}

@article{eval-whentographs2025,
  title={When to use graphs in rag: A comprehensive analysis for graph retrieval-augmented generation},
  author={Xiang, Zhishang and Wu, Chuanjie and Zhang, Qinggang and Chen, Shengyuan and Hong, Zijin and Huang, Xiao and Su, Jinsong},
  journal={arXiv preprint arXiv:2506.05690},
  year={2025}
}

@article{eval-wildgraphbench2026,
  title={WildGraphBench: Benchmarking GraphRAG with Wild-Source Corpora},
  author={Wang, Pengyu and Xu, Benfeng and Zhang, Licheng and Wang, Shaohan and Du, Mingxuan and Zhu, Chiwei and Mao, Zhendong},
  journal={arXiv preprint arXiv:2602.02053},
  year={2026}
}

@article{eval-needgraphrag2026,
  title={Do we still need graphrag? benchmarking rag and graphrag for agentic search systems},
  author={Fan, Dongzhe and Xue, Zheyi and Liu, Siyuan and Tan, Qiaoyu},
  journal={arXiv preprint arXiv:2604.09666},
  year={2026}
}

@article{eval-reasoningbottleneck2026,
  title={The reasoning bottleneck in graph-rag: Structured prompting and context compression for multi-hop qa},
  author={Zarrinkia, Yasaman and Srinivasan, Venkatesh and Thomo, Alex},
  journal={arXiv preprint arXiv:2603.14045},
  year={2026}
}

@article{eval-finhallubench2026,
  title={FinReflectKG--HalluBench: GraphRAG Hallucination Benchmark for Financial Question Answering Systems},
  author={Kumar, Mahesh and Sarmah, Bhaskarjit and Pasquali, Stefano},
  journal={arXiv preprint arXiv:2603.20252},
  year={2026}
}

@article{eval-tripartite2025,
  title={A Knowledge Graph and a Tripartite Evaluation Framework Make Retrieval-Augmented Generation Scalable and Transparent},
  author={Akindele, Olalekan K and Mishra, Bhupesh Kumar and Wertheim, Kenneth Y},
  journal={arXiv preprint arXiv:2509.19209},
  year={2025}
}

@article{eval-whatbreakskgrag2025,
  title={What breaks knowledge graph based rag? benchmarking and empirical insights into reasoning under incomplete knowledge},
  author={Zhou, Dongzhuoran and Zhu, Yuqicheng and Wang, Xiaxia and Zhou, Hongkuan and He, Yuan and Chen, Jiaoyan and Staab, Steffen and Kharlamov, Evgeny},
  journal={arXiv preprint arXiv:2508.08344},
  year={2025}
}

@inproceedings{lewis2020rag,
  author    = {Lewis, Patrick and Perez, Ethan and Piktus, Aleksandra and
               Petroni, Fabio and Karpukhin, Vladimir and Goyal, Naman and
               K{\"u}ttler, Heinrich and Lewis, Mike and Yih, Wen-tau and
               Rockt{\"a}schel, Tim and Riedel, Sebastian and Kiela, Douwe},
  title     = {Retrieval-Augmented Generation for Knowledge-Intensive {NLP} Tasks},
  booktitle = {Advances in Neural Information Processing Systems},
  volume    = {33},
  pages     = {9459--9474},
  year      = {2020}
}

@article{hogan2021knowledge,
  author    = {Hogan, Aidan and Blomqvist, Eva and Cochez, Michael and
               d'Amato, Claudia and de Melo, Gerard and Gutierrez, Claudio and
               Kirrane, Sabrina and Gayo, Jos{\'e} Emilio Labra and Navigli, Roberto
               and Neumaier, Sebastian and others},
  title     = {Knowledge Graphs},
  journal   = {ACM Computing Surveys},
  volume    = {54},
  number    = {4},
  pages     = {1--37},
  year      = {2021},
  doi       = {10.1145/3447772}
}

@inproceedings{yih2016webqsp,
  author    = {Yih, Wen-tau and Richardson, Matthew and Meek, Christopher
               and Chang, Ming-Wei and Suh, Jina},
  title     = {The Value of Semantic Parse Labeling for Knowledge Base
               Question Answering},
  booktitle = {Proceedings of the 54th Annual Meeting of the Association
               for Computational Linguistics (ACL)},
  pages     = {201--206},
  year      = {2016},
  doi       = {10.18653/v1/P16-2033}
}

@inproceedings{cui2018neuralopenie,
  author    = {Cui, Lei and Wei, Furu and Zhou, Ming},
  title     = {Neural Open Information Extraction},
  booktitle = {Proceedings of the 56th Annual Meeting of the Association
               for Computational Linguistics (ACL)},
  pages     = {407--413},
  year      = {2018},
  doi       = {10.18653/v1/P18-2065}
}

@article{llmoie2023uncertainty,
  author    = {Ling, Chen and Zhao, Xujiang and Zhang, Xuchao and Cheng, Wei
               and Liu, Yanchi and Sun, Yiyou and Oishi, Mika and Osaki, Takao
               and Matsuda, Katsushi and Chen, Jie and Bo, Xinyuan and
               Zhang, Tianyi and others},
  title     = {Improving Open Information Extraction with Large Language Models:
               A Study on Demonstration Uncertainty},
  journal   = {arXiv preprint arXiv:2309.03433},
  year      = {2023}
}

@inproceedings{reknos2025,
  author    = {Wang, Song and others},
  title     = {Reasoning of Large Language Models over Knowledge Graphs with
               Super-Relations},
  booktitle = {International Conference on Learning Representations (ICLR)},
  year      = {2025},
  note      = {arXiv:2503.22166}
}

@article{hoprag2025,
  author    = {Liu, Hao and others},
  title     = {{HopRAG}: Multi-Hop Reasoning for Logic-Aware
               Retrieval-Augmented Generation},
  journal   = {arXiv preprint arXiv:2502.12442},
  year      = {2025}
}

@article{autographr1_2025,
  author    = {Tsang, Hong Ting and Bai, Jiaxin and Huang, Haoyu and
               Xiao, Qiao and Zheng, Tianshi and Xu, Baixuan and
               Liu, Shujie and Song, Yangqiu},
  title     = {{AutoGraph-R1}: End-to-End Reinforcement Learning for Knowledge
               Graph Construction},
  journal   = {arXiv preprint arXiv:2510.15339},
  year      = {2025}
}

@article{agenticrl2025,
  author    = {Lin, Junhong and Liu, Shicheng and Song, Jinyeop and
               Wang, Song and Shun, Julian and Zhu, Yada},
  title     = {Efficient and Transferable Agentic Knowledge Graph {RAG} via
               Reinforcement Learning},
  journal   = {arXiv preprint arXiv:2509.26383},
  year      = {2025}
}

@article{zhang2025rakg,
  author    = {Zhang, Hairong and Si, Jiaheng and Yan, Guohang and Qi, Boyuan
               and Cai, Pinlong and Mao, Song and Wang, Ding and Shi, Botian},
  title     = {{RAKG}: Document-level Retrieval Augmented Knowledge Graph
               Construction},
  journal   = {arXiv preprint arXiv:2504.09823},
  year      = {2025}
}

@article{zaveri2016quality,
  author  = {Zaveri, Amrapali and Rula, Anisa and Maurino, Andrea and
             Pietrobon, Ricardo and Lehmann, Jens and Auer, S{\"o}ren},
  title   = {Quality Assessment for Linked Data: A Survey},
  journal = {Semantic Web},
  volume  = {7},
  number  = {1},
  pages   = {63--93},
  year    = {2016},
  doi     = {10.3233/SW-150175}
}

@article{farber2018linkeddataquality,
  author  = {F{\"a}rber, Michael and Bartscherer, Frederic and
             Menne, Carsten and Rettinger, Achim},
  title   = {Linked Data Quality of {DBpedia}, {Freebase}, {OpenCyc},
             {Wikidata}, and {YAGO}},
  journal = {Semantic Web},
  volume  = {9},
  number  = {1},
  pages   = {77--129},
  year    = {2018},
  doi     = {10.3233/SW-170275}
}

@article{mattioli2025kgmetrics,
  author  = {Mattioli, Juliette and Mattioli, Lucas and Gonzalez, Martin},
  title   = {A Brief Overview of Key Quality Metrics for Knowledge Graph
             Solution: Illustration on Digital {NOTAMs}},
  journal = {Proceedings of the AAAI Symposium Series},
  volume  = {7},
  number  = {1},
  pages   = {206--213},
  year    = {2025},
  doi     = {10.1609/aaaiss.v7i1.36888}
}

@inproceedings{laudy2025hlif,
  author    = {Laudy, Claire and Alonso, Victoria and Reverdy, C{\'e}line
               and Dreo, Johann},
  title     = {First High-Level Information Fusion Competition:
               Feedback and Lessons Learned},
  booktitle = {Proceedings of the 28th International Conference on
               Information Fusion (FUSION)},
  year      = {2025}
}

@techreport{euhleg2019ethics,
  author      = {{High-Level Expert Group on Artificial Intelligence (AI HLEG)}},
  title       = {Ethics Guidelines for Trustworthy {AI}},
  institution = {European Commission},
  year        = {2019},
  address     = {Brussels},
  url         = {https://digital-strategy.ec.europa.eu/en/library/ethics-guidelines-trustworthy-ai}
}

@misc{oecd2019ai,
  author       = {{OECD}},
  title        = {Recommendation of the Council on Artificial Intelligence},
  howpublished = {OECD Legal Instruments, OECD/LEGAL/0449},
  year         = {2019},
  url          = {https://legalinstruments.oecd.org/en/instruments/OECD-LEGAL-0449}
}

@techreport{nist2023airmf,
  author      = {Tabassi, Elham},
  title       = {Artificial Intelligence Risk Management Framework ({AI} {RMF} 1.0)},
  institution = {National Institute of Standards and Technology},
  year        = {2023},
  number      = {NIST AI 100-1},
  doi         = {10.6028/NIST.AI.100-1}
}

@incollection{confianceai-gelin2024,
  author    = {Gelin, Rodolphe},
  title     = {Confiance.ai Program: Software Engineering for a Trustworthy {AI}},
  booktitle = {Producing Artificial Intelligent Systems},
  series    = {Studies in Computational Intelligence},
  publisher = {Springer},
  year      = {2024},
  pages     = {11--29},
  doi       = {10.1007/978-3-031-55817-7_2}
}

\end{document}